\RequirePackage{fix-cm}
\documentclass[aps,prd,10pt,twocolumn,amsmath,superscriptaddress,amssymb,preprintnumbers,showpacs,floatfix,nofootinbib,longbibliography]{revtex4-2}

\pdfoutput=1
\usepackage{hyperref}
\usepackage{graphicx}
\usepackage{appendix}
\usepackage{bm}
\usepackage{amsfonts,amsmath,amssymb,bm,bbm,mathrsfs}
\usepackage{color}
\usepackage{url}
\usepackage{siunitx}
\usepackage[english]{babel}
\usepackage[capitalize,nameinlink]{cleveref}
\crefformat{equation}{Eq.~(#2#1#3)}
\Crefname{section}{Sec.}{Secs.}
\Crefformat{section}{Sec.~#2#1#3}
\Crefname{equation}{Eq.}{Eqs.}
\Crefname{figure}{Fig.}{Figs.}
\Crefformat{figure}{Fig.~#2#1#3}
\Crefmultiformat{figure}
  {Fig.~#2#1#3}
  { and~#2#1#3}
  {, #2#1#3}
  {, and~#2#1#3}
\Crefformat{table}{Table~#2#1#3}
\usepackage{float}
\usepackage{slashed}
\usepackage{enumitem}
\usepackage{leftindex}
\usepackage[compat=1.1.0]{tikz-feynman}
\usepackage{feynmp-auto}

\addto\extrasenglish{%
}
\addto\extrasenglish{%
}

\hypersetup{
    pdfnewwindow=orange,      % links in new window
    colorlinks=true,       % false: boxed links; true: colored links
    linkcolor=orange,          % color of internal links
    citecolor=orange,        % color of links to bibliography
    filecolor=orange,      % color of file links
    urlcolor=orange        % color of external links
}

\usepackage{lettrine}
\usepackage[dvipsnames]{xcolor}

\input Zallman.fd

\LettrineTextFont{\itshape}
\newcommand{\dd}{\mathrm{d}}
\newcommand{\appropto}{\mathrel{\vcenter{
  \offinterlineskip\halign{\hfil$##$\cr
    \propto\cr\noalign{\kern2pt}\sim\cr\noalign{\kern-2pt}}}}}

\DeclareMathSizes{10}{10}{5}{5}

\begin{document}

\preprint{SLAC-PUB-260709}

\title{Dark Matter Weather: \\ Probing Sub-GeV Interactions with Earth-Shielding Modulation}

\author{Tetiana Kozynets} \thanks{\href{mailto:tetiana.kozynets@liverpool.ac.uk}{tetiana.kozynets@liverpool.ac.uk}; \href{http://orcid.org/0000-0003-2723-3931}{0000-0003-2723-3931}}
\affiliation{Department of Mathematical Sciences, 
University of Liverpool, Liverpool, L69 7ZL, United Kingdom}

\author{Rebecca K. Leane}
\thanks{\href{mailto:rleane@stanford.edu}{rleane@stanford.edu}; \href{http://orcid.org/0000-0002-1287-8780}{0000-0002-1287-8780}}
\affiliation{SLAC National Accelerator Laboratory, Stanford University, Menlo Park, CA 94025, USA}
\affiliation{Kavli Institute for Particle Astrophysics and Cosmology, Stanford University, Stanford, CA 94305, USA}

\author{Juri Smirnov}\thanks{\href{mailto:juri.smirnov@liverpool.ac.uk}{juri.smirnov@liverpool.ac.uk}; \href{ http://orcid.org/0000-0002-3082-0929}{0000-0002-3082-0929}}
\affiliation{Department of Mathematical Sciences, 
University of Liverpool, Liverpool, L69 7ZL, United Kingdom}

\date{\today}

%%%%%%%%%%%%%%%%%%%%%%%%%%%%%%%%%%%%%%%%%%%%%%%%%%%%%%%%%
%%%%%%%%%%%%%%%%%%%%%%%%%%%%%%%%%%%%%%%%%%%%%%%%%%%%%%%%%

\begin{abstract}
Daily modulation from Earth shielding provides a powerful search handle for sub-GeV dark matter (DM) in low-threshold experiments. We use this effect as a probe of interaction structure in scenarios where DM couples to both electrons and nuclei. Nuclear scattering in the Earth alters the incident DM flux, while electron scattering produces the observable ionization signal, so the total rate and modulation pattern encode different aspects of the underlying interactions. We develop this two-interaction framework for argon and xenon targets and introduce a new statistical analysis that tests the modulation shape, including the location-dependent exposure of underground detectors to different Earth-crossing trajectories, alone and in combination with the ionization spectrum. We demonstrate the method for liquid-noble detectors at underground sites SURF (USA), LNGS (Italy), and SUPL (Australia), and apply it to DarkSide-50 data as a concrete case study. Our results show that Earth-scattering modulation can help disentangle electron and nuclear DM interactions and provide a validation handle for low-threshold liquid-noble searches.
\end{abstract}

\maketitle

\begin{figure}[t]
\includegraphics[width=\columnwidth]{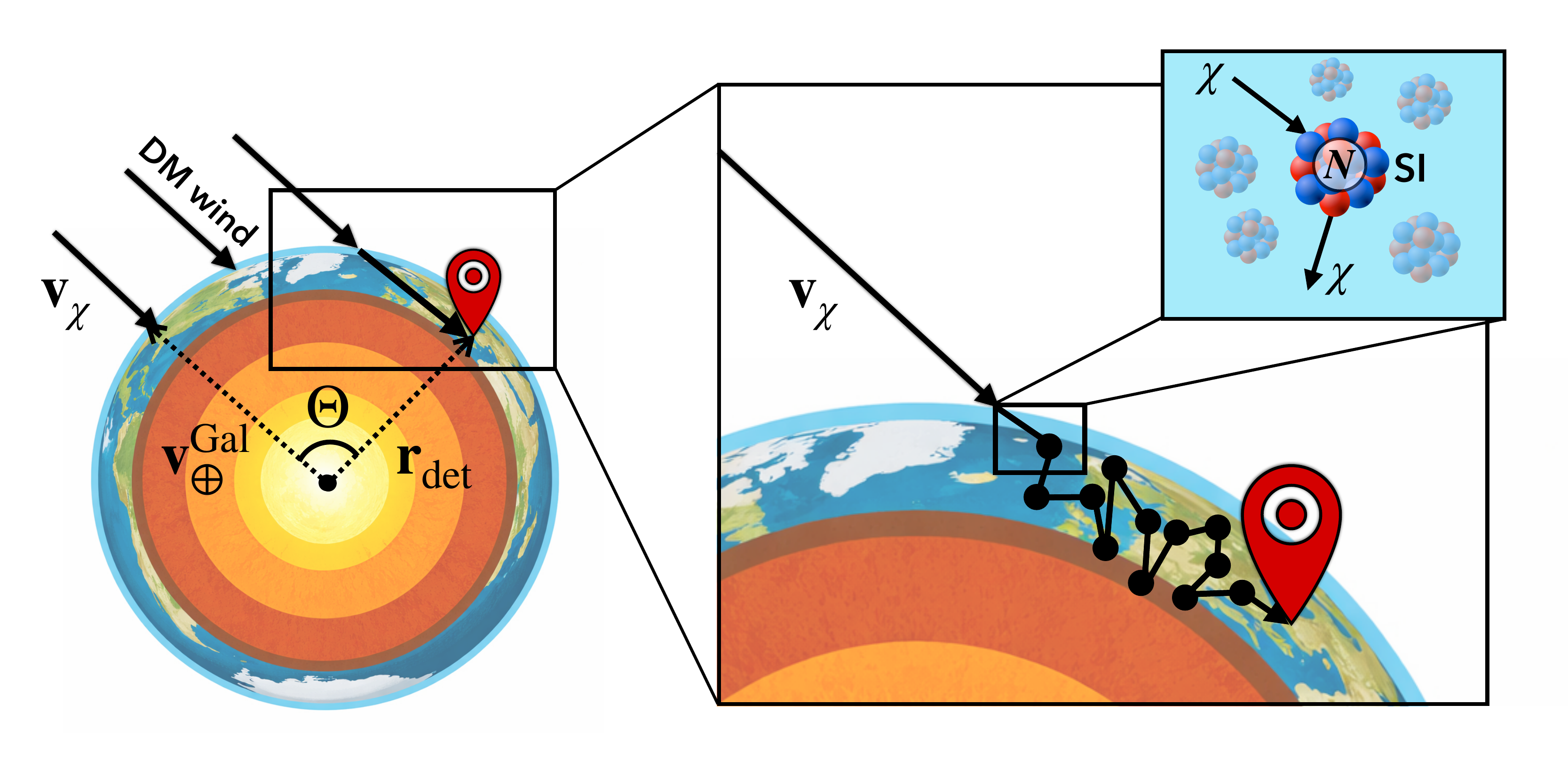}
\caption{Illustration of spin-independent DM--nucleus scattering in the Earth en route to a detector located at $\mathbf{r}_{\mathrm{det}}$.}
\vspace{-4mm}
\label{fig:earth_diffusion}
\end{figure}

\lettrine{H}{ow is the dark matter weather today?} In ordinary direct-detection searches, the answer is assumed to be nearly the same all day long: a steady Galactic wind, with small annual variations from the Earth’s orbit around the Sun. For dark matter (DM) that scatters sufficiently in the Earth, however, the forecast can acquire a daily cycle. Much like California’s \textit{Standard Model} weather, tomorrow will generally resemble today, but the local conditions still change from morning to afternoon to evening. As the Earth rotates each day, an underground detector samples DM arriving along different terrestrial path lengths, so the Earth itself can attenuate, deflect, and reshape the incident flux before it reaches the target; the mechanism is schematically shown in Fig.~\ref{fig:earth_diffusion}. The result is a predictable daily pattern whose phase and amplitude are tied to the detector location and the DM interaction strength~\cite{Collar:1992qc,Collar:1993ss,Hasenbalg:1997hs}. 

Such temporal structure is especially valuable for sub-GeV DM, a theoretically well-motivated target that is experimentally difficult to test~\cite{Essig:2022dfa,Alfonso-Pita:2022akn,Krnjaic:2022ozp,SuperCDMS:2022kse,Akesson:2022vza,Mitridate:2022tnv,Wang:2022cyk,Battaglieri:2017aum}. In this mass range, conventional nuclear-recoil searches rapidly lose sensitivity, while electron recoils and other low-threshold observables provide powerful discovery channels~\cite{Essig:2011nj,Essig:2022dfa,Alfonso-Pita:2022akn,Krnjaic:2022ozp,SuperCDMS:2022kse,Akesson:2022vza,Mitridate:2022tnv,Wang:2022cyk,Battaglieri:2017aum,Graham:2012su,Essig:2015cda,Hochberg:2015pha,Hochberg:2015fth,Hochberg:2019cyy,Derenzo:2016fse,Schutz:2016tid,Hochberg:2016ntt,Essig:2016crl,Hochberg:2017wce,Cavoto:2017otc,Emken:2017erx,Emken:2017qmp,Griffin:2018bjn,Sanchez-Martinez:2019bac,Essig:2019xkx,Emken:2019tni,Kurinsky:2019pgb,Geilhufe:2019ndy,Blanco:2019lrf,Baxter:2019pnz,Catena:2019gfa,Radick:2020qip,Gelmini:2020xir,Trickle:2020oki,Kurinsky:2020dpb,Griffin:2020lgd,Blanco:2021hlm,Knapen:2021run,Hochberg:2021ymx,Hochberg:2021yud,Essig:2022dfa,Hochberg:2022apz,Das:2022srn,Das:2024jdz,Griffin:2024cew,Cook:2024cgm,Simchony:2024kcn,QROCODILE:2024zmg,Blanco:2022cel,Hochberg:2025rjs,Berghaus:2026kmj,Dreyer:2026bmz,Abbamonte:2025guf,Leane:2025efj,Santos-Olmsted:2025nuk}. Yet these same low-energy channels are often accompanied by poorly understood detector backgrounds, including spurious electrons, dark counts, instrumental noise, and low-level radioactivity. A robust discovery claim is therefore strengthened by distinctive structure in time, energy, target material, and detector location.

Time-dependent signals have long been recognized as powerful diagnostics in direct detection. Annual modulation due to the Earth’s orbit around the Sun is the classic example~\cite{Drukier:1986tm}, with additional time-dependent effects arising from gravitational focusing~\cite{Sikivie:2002bj,Alenazi:2006wu} and the Earth’s rotation~\cite{Kouvaris:2015xga}. Earth-shielding modulation, discussed above, has been developed with dedicated simulation and semi-analytic tools~\cite{Emken:2017hnp,Kavanagh:2016pyr,Emken:2017qmp,Kavanagh:2017cru,Emken:2018run,Emken:2019tni,Cappiello:2023hza}, applied to sub-GeV electron-recoil searches~\cite{Bertou:2025adb}, and recently searched for in low-threshold data~\cite{SENSEI:2025qvp,DAMIC-M:2025ltz}. In this work, we use Earth-shielding modulation for a complementary purpose: to probe the interaction structure of DM.

We focus on scenarios in which DM couples to both electrons and nuclei. Such models arise naturally in many simple extensions of the Standard Model (see Appendix~\ref{app:models}). In single-mediator models, such as a heavy dark photon, electron and nuclear scattering are correlated because both arise from the same underlying interaction. In multi-mediator models, such as theories with separate baryon and lepton gauge interactions, the two couplings can instead vary more independently. These possibilities motivate treating the DM--electron and DM--nucleon scattering cross sections as two related but distinct quantities. Earth-scattering modulation is particularly well suited to this problem: nuclear scattering in the Earth controls the time-dependent attenuation of the incident flux, while electron scattering in the detector produces the observable ionization signal. The total rate and the modulation pattern therefore encode complementary information.

We develop this two-interaction framework for liquid-argon and liquid-xenon detectors, quantifying how the daily modulation depends on DM mass, detector depth and location, target material, and the two scattering cross sections. Detectors in different hemispheres can sample different shielding regimes over the course of a day, making site dependence an important part of the signal interpretation.  We also introduce a statistical analysis that uses the modulation shape, alone and in combination with the ionization spectrum. This is essential because the modulation pattern can carry information that is not present in a rate-only search. We demonstrate the method for representative underground sites and for DarkSide-50 data as a concrete case study. Although public DarkSide-50 data do not include event timing information, the measured ionization spectrum and the assumption of no observed daily modulation allow us to illustrate how existing data can constrain the combined electron- and nuclear-scattering parameter space.

The paper is organized as follows. In Sec.~\ref{sec:shielding}, we describe the modification of the DM velocity distribution and DM density by Earth scattering. In Sec.~\ref{sec:modulation}, we compute the resulting daily modulation in liquid-noble ionization signals. In Sec.~\ref{sec:stat_analysis}, we introduce the statistical framework used to test the modulation strength and discriminate the modulated DM signals against time-independent backgrounds. In Sec.~\ref{sec:results_general_nobles}, we present sensitivity projections for argon and xenon detectors at representative underground sites. In Sec.~\ref{sec:darkside_case}, we apply the framework to DarkSide-50 as a case study and show how to combine it with spectral information. We conclude by discussing the role of Earth-scattering modulation as both a discovery channel and a validation tool for sub-GeV DM searches.

\section{Earth Shielding Effects}
\label{sec:shielding}

\subsection{Unattenuated Detector-Frame DM Distribution}
\label{sec:unattenuated_distribution}

Before entering the Earth, Galactic DM is assumed to follow the velocity distribution of an isotropic, isothermal halo. In the Galactic frame, this is described by the truncated Maxwell--Boltzmann distribution~\cite{Baxter:2021pqo}
\begin{equation}
f^{(3)}(\mathbf{v}_{\chi}^{\mathrm{Gal}})
=
\frac{1}{\mathcal{N}}
\exp\left(
-\frac{|\mathbf{v}_{\chi}^{\mathrm{Gal}}|^2}{2\sigma_0^2}
\right)
\mathbb{H}\left(
v_{\mathrm{esc}}-|\mathbf{v}_{\chi}^{\mathrm{Gal}}|
\right),
\label{eq:maxwell_boltzmann_galactic_dist}
\end{equation}
where $\sigma_0=v_0/\sqrt{2}$ is the DM velocity dispersion, with $v_0=238\,\mathrm{km\,s^{-1}}$~\cite{Bland:2016xxx,GRAVITY:2021xxx}, $v_{\mathrm{esc}}=544\,\mathrm{km\,s^{-1}}$ is the Galactic escape velocity~\cite{Smith:2006ym}, and $\mathbb{H}$ is the Heaviside step function. The normalization factor is~\cite{QUEST-DMC:2025qsa}
\begin{align}
\mathcal{N}
=
\Bigg[
\mathrm{erf}\left(\frac{v_{\mathrm{esc}}}{v_0}\right)
-
\frac{2}{\sqrt{\pi}}
\frac{v_{\mathrm{esc}}}{v_0}
\exp\left(-\frac{v_{\mathrm{esc}}^2}{v_0^2}\right)
\Bigg]
(\sqrt{\pi}v_0)^3.
\end{align}

In the absence of in-Earth scattering, the DM velocity in the detector frame is
\begin{equation}
\mathbf{v}
\equiv
\mathbf{v}_{\chi}^{\mathrm{det}}
=
\mathbf{v}_{\chi}^{\mathrm{Gal}}
-
\mathbf{v}_{\mathrm{det}}^{\mathrm{Gal}},
\label{eq:dm_vel_detector_frame}
\end{equation}
where $\mathbf{v}_{\mathrm{det}}^{\mathrm{Gal}}$ is the detector velocity in the Galactic frame. Combining the local circular velocity,
$\mathbf{v}_0=(0,238,0)\,\mathrm{km\,s^{-1}}$,
the Sun's peculiar velocity,
$\mathbf{v}_{\odot}=(11.1,12.2,7.3)\,\mathrm{km\,s^{-1}}$~\cite{Schonrich:2010xxx},
and the Earth's orbital velocity about the Sun,
$\mathbf{v}_{\oplus}^{\odot}(t)$,
gives the velocity of the Earth in the Galactic frame,
\begin{equation}
\mathbf{v}_{\oplus}^{\mathrm{Gal}}(t)
=
\mathbf{v}_0
+
\mathbf{v}_{\odot}
+
\mathbf{v}_{\oplus}^{\odot}(t).
\label{eq:earth_vel_gal_frame}
\end{equation}
The detector velocity is then
\begin{equation}
\mathbf{v}_{\mathrm{det}}^{\mathrm{Gal}}(t)
=
\mathbf{v}_{\oplus}^{\mathrm{Gal}}(t)
+
\mathbf{v}_{\mathrm{det}}^{\oplus}(t),
\label{eq:detector_vel_galactic_frame}
\end{equation}
where $\mathbf{v}_{\mathrm{det}}^{\oplus}(t)$ denotes the contribution due to the Earth's axial rotation. Since
$|\mathbf{v}_{\mathrm{det}}^{\oplus}|
\ll
|\mathbf{v}_{\oplus}^{\mathrm{Gal}}|$,
we take
$\mathbf{v}_{\mathrm{det}}^{\mathrm{Gal}}
\simeq
\mathbf{v}_{\oplus}^{\mathrm{Gal}}$.

Defining the apparent DM-wind velocity as
$\mathbf{v}_{\chi}\equiv-\mathbf{v}_{\oplus}^{\mathrm{Gal}}$,
the unattenuated detector-frame distribution is
\begin{equation}
f^{(3)}(\mathbf{v})
=
\frac{1}{\mathcal{N}}
\exp\left(
-\frac{|\mathbf{v}-\mathbf{v}_{\chi}|^2}{2\sigma_0^2}
\right)
\mathbb{H}\left(
v_{\mathrm{esc}}-|\mathbf{v}-\mathbf{v}_{\chi}|
\right).
\label{eq:unattenuated_shm_final_expression}
\end{equation}
We refer to \Cref{eq:unattenuated_shm_final_expression} as the Standard Halo Model (SHM)~\cite{Baxter:2021pqo}. Its corresponding speed distribution is
\begin{equation}
f(v)
\equiv
f^{(1)}(v)
=
\int_0^{2\pi}
\int_0^\pi
v^2 f^{(3)}(\mathbf{v})
\sin\theta\,
\mathrm{d}\theta\,
\mathrm{d}\phi\,
\label{eq:shm_1d}
\end{equation}
where
$\mathbf{v}
=
(v\sin\theta\cos\phi,
v\sin\theta\sin\phi,
v\cos\theta)$.

For the DM density, we take the unattenuated local DM density to be $\rho=0.3~$GeV/cm$^3$.

\begin{figure}[t]
\includegraphics[width=\columnwidth]{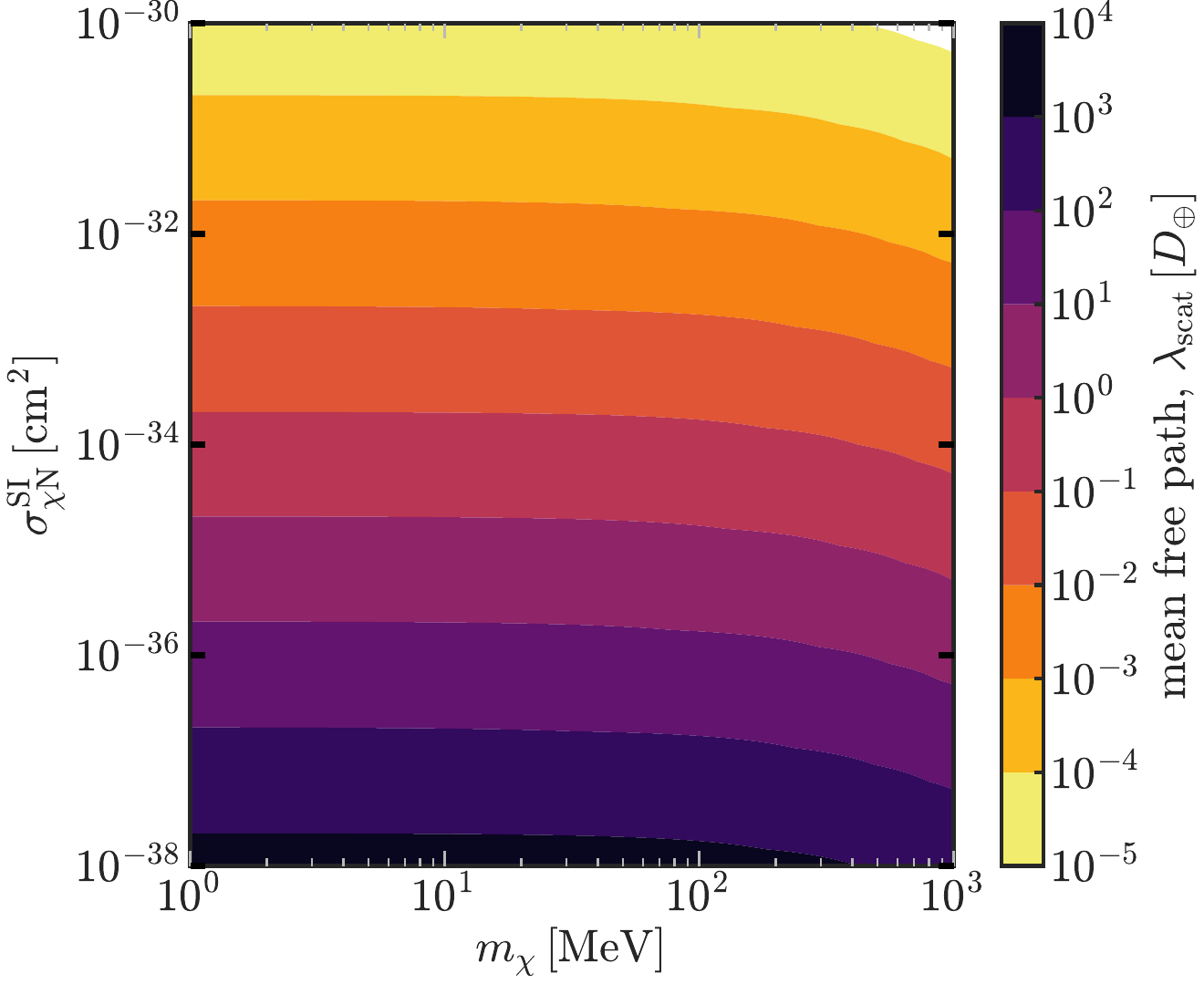}
\caption{Mean free path $\lambda_{\mathrm{scat}}$ for DM--nucleus scattering inside the Earth, shown in units of the Earth's diameter $D_{\oplus}$ for a contact DM--nucleon interaction.}
\label{fig:mean_free_path}
\end{figure}

\begin{figure}[H]
\includegraphics[width=\columnwidth]{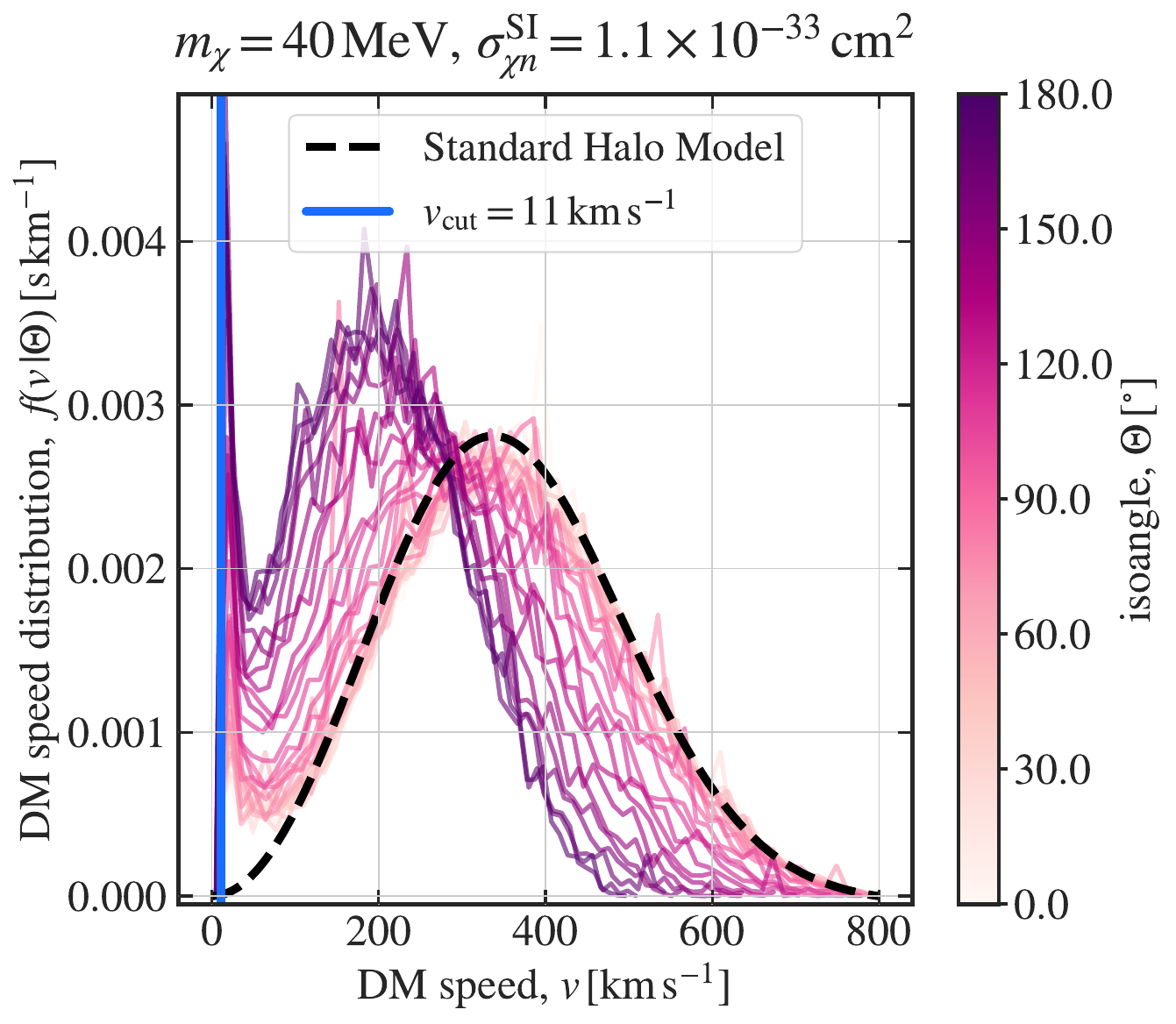}
\caption{Representative Earth-shielded DM speed distributions as a function of the isoangle $\Theta$. The dashed black line shows the unattenuated SHM distribution from \Cref{eq:shm_1d}, while the color-coded lines are obtained from \texttt{DaMaSCUS} simulations~\cite{Emken:2017qmp} for a detector depth $d_{\mathrm{det}}=1400\mathrm{m}$. The solid blue line denotes the minimum simulated speed, corresponding to the Earth's escape velocity. The displayed DM mass and cross section provide a representative example from the simulation grid.}
\label{fig:attenuated_velocities}
\end{figure}

\subsection{Earth-Scattering Scale and Shielding Geometry}
\label{sec:shielding_geometry}

The importance of Earth shielding is controlled by two quantities: the DM mean free path in the Earth and the distance travelled before reaching the detector. 

Figure~\ref{fig:mean_free_path} shows the mean free path $\lambda_{\mathrm{scat}}$ for spin-independent DM--nucleus scattering through a contact interaction, assuming the average Earth composition of Ref.~\cite{Emken:2018run}. We find that
$\lambda_{\mathrm{scat}}\lesssim D_{\oplus}$
for
$\sigma_{\chi n}^{\mathrm{SI}}\gtrsim10^{-34}\,\mathrm{cm^2}$,
where $D_{\oplus}=2R_{\oplus}$ is the Earth's diameter. A DM particle traversing one Earth diameter therefore undergoes approximately one scatter on average at this cross section, with larger cross sections producing more frequent scattering. Throughout this work, we restrict our attention to the heavy-mediator scenario. For light mediators, atomic screening suppresses scattering rates at low momentum transfers, while the remaining interactions are forward-peaked and inefficient at changing DM energy and direction. Thus, comparable daily modulation requires substantially larger reference cross sections than in the contact case; see Appendix~\ref{app:light_mediator_modulation}.

For a fixed DM--nucleon cross section, the degree of shielding depends on the propagation distance through the Earth. We parameterize this distance using the angle between the detector position vector $\mathbf{r}_{\mathrm{det}}$ and the direction from which the apparent DM wind arrives~\cite{Emken:2017qmp,Bertou:2025adb},
\begin{equation}
\Theta(t)
=
\cos^{-1}\left[
\frac{
\mathbf{v}_{\oplus}^{\mathrm{Gal}}(t)
\cdot
\mathbf{r}_{\mathrm{det}}(t)
}{
v_{\oplus}^{\mathrm{Gal}}(t)
(R_{\oplus}-d_{\mathrm{det}})
}
\right],
\label{eq:isoangle_definition}
\end{equation}
where $d_{\mathrm{det}}$ is the detector depth,
$|\mathbf{r}_{\mathrm{det}}|=R_{\oplus}-d_{\mathrm{det}}$,
and $\mathbf{r}_{\mathrm{det}}(t)$ varies with the Earth's sidereal rotation. The angle $\Theta$ is also called the isodetection angle, or \textit{isoangle}, because detectors at the same value of $\Theta$ experience the same attenuated DM flux.

When $\Theta\simeq0^\circ$, the Earth's Galactic velocity is aligned with the detector position vector, so the average DM wind arrives from above and particles typically traverse only the local overburden, of order $d_{\mathrm{det}}$. By contrast, $\Theta\simeq180^\circ$ corresponds to DM arriving through nearly the full Earth. Larger $\Theta$ therefore generally implies a longer propagation distance, more scattering, and stronger Earth shielding.

\begin{figure}[t]
\includegraphics[width=\columnwidth]{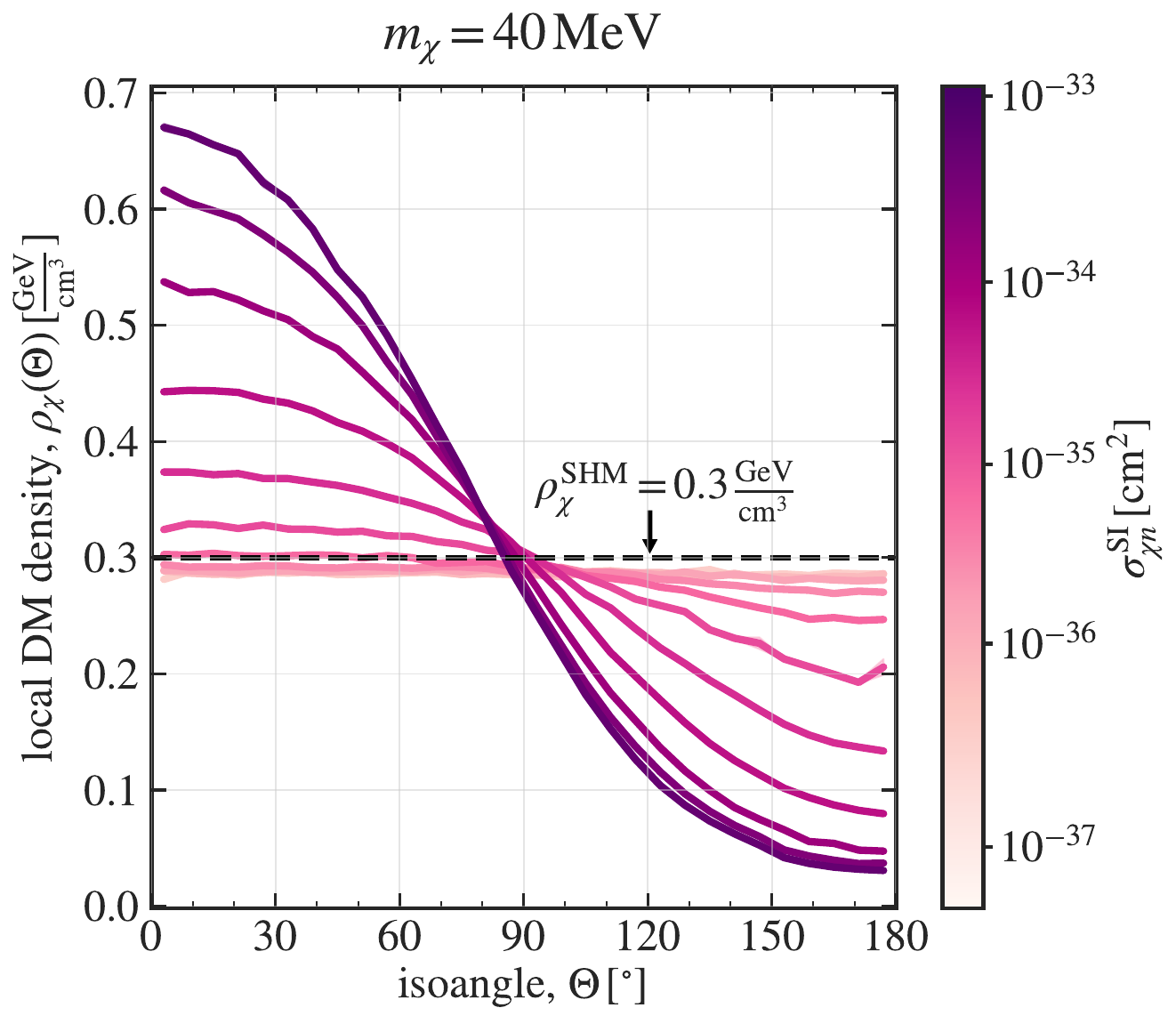}
\caption{Local DM density $\rho_{\chi}$ as a function of the isoangle $\Theta$, obtained from \texttt{DaMaSCUS} simulations for a range of spin-independent DM--nucleon cross sections $\sigma_{\chi n}^{\mathrm{SI}}$. The choice $m_{\chi}=40\,\mathrm{MeV}$ provides a representative example.}
\label{fig:local_dm_density_vs_isoangle}
\end{figure}

\subsection{Earth-Modified DM Speed Distribution and Density}
\label{sec:modified_distributions}

DM--nucleus scattering modifies both the speed distribution and the local density of DM incident on the detector. Individual collisions reduce the DM kinetic energy and shift the speed distribution toward lower values, while deflections alter the flux reaching a given detector location~\cite{Cappiello:2023hza,QUEST-DMC:2025qsa,Emken:2017qmp,Bertou:2025adb}. We calculate these effects using Monte Carlo simulations with the public \texttt{DaMaSCUS} package~\cite{Emken:2017qmp,DaMaSCUS:2020xxx}.

Figure~\ref{fig:attenuated_velocities} shows representative Earth-shielded speed distributions as a function of $\Theta$. Relative to the unattenuated SHM distribution, scattering shifts the DM population toward lower speeds and suppresses the high-speed tail, with stronger modifications at larger $\Theta$. We denote the resulting distributions by $f(v|\Theta)$ to emphasize their isoangle dependence.

Figure~\ref{fig:local_dm_density_vs_isoangle} illustrates how intraterrestrial scattering also introduces an isoangle dependence in the local DM density $\rho_{\chi}$. Energy loss and deflection deplete the incident DM flux at large $\Theta$ and partially redistribute it toward smaller $\Theta$~\cite{Emken:2017qmp}.

The modified speed distribution $f(v|\Theta)$ and local density $\rho_{\chi}(\Theta)$ together determine the isoangle dependence of the in-detector ionization rates calculated in the upcoming \Cref{sec:scat_rate_calculation}.

\section{Daily Modulation Signal}\label{sec:modulation}

\subsection{Isoangle Time Dependence}\label{sec:isoangle_time_dependence}

The time dependence of the isoangle $\Theta(t)$ defined in \Cref{eq:isoangle_definition} arises from:
\begin{itemize}
\item $\mathbf{r}_{\mathrm{det}}(t)$, whose sidereal rotation drives the daily modulation studied here;
\item $\mathbf{v}_{\oplus}^{\odot}(t)$, whose annual variation changes $\mathbf{v}_{\oplus}^{\mathrm{Gal}}(t)$ through \Cref{eq:earth_vel_gal_frame}. We neglect this annual time dependence and fix the orbital velocity to its value on March~9, a representative choice that approximately reproduces the annually averaged detector-frame speed distribution in standard nondirectional calculations~\cite{Baxter:2021pqo}.
Thus, we take
\begin{equation}
\mathbf{v}_{\oplus}^{\odot}(t)
\longrightarrow
\mathbf{v}_{\oplus}^{\odot}(\text{March 9})
\simeq
(29.5,0.0,5.8)\ {\rm km\,s^{-1}}.
\label{eq:fixed_orbital_velocity}
\end{equation}
\end{itemize}

\begin{figure}[t]
\includegraphics[width=\columnwidth]{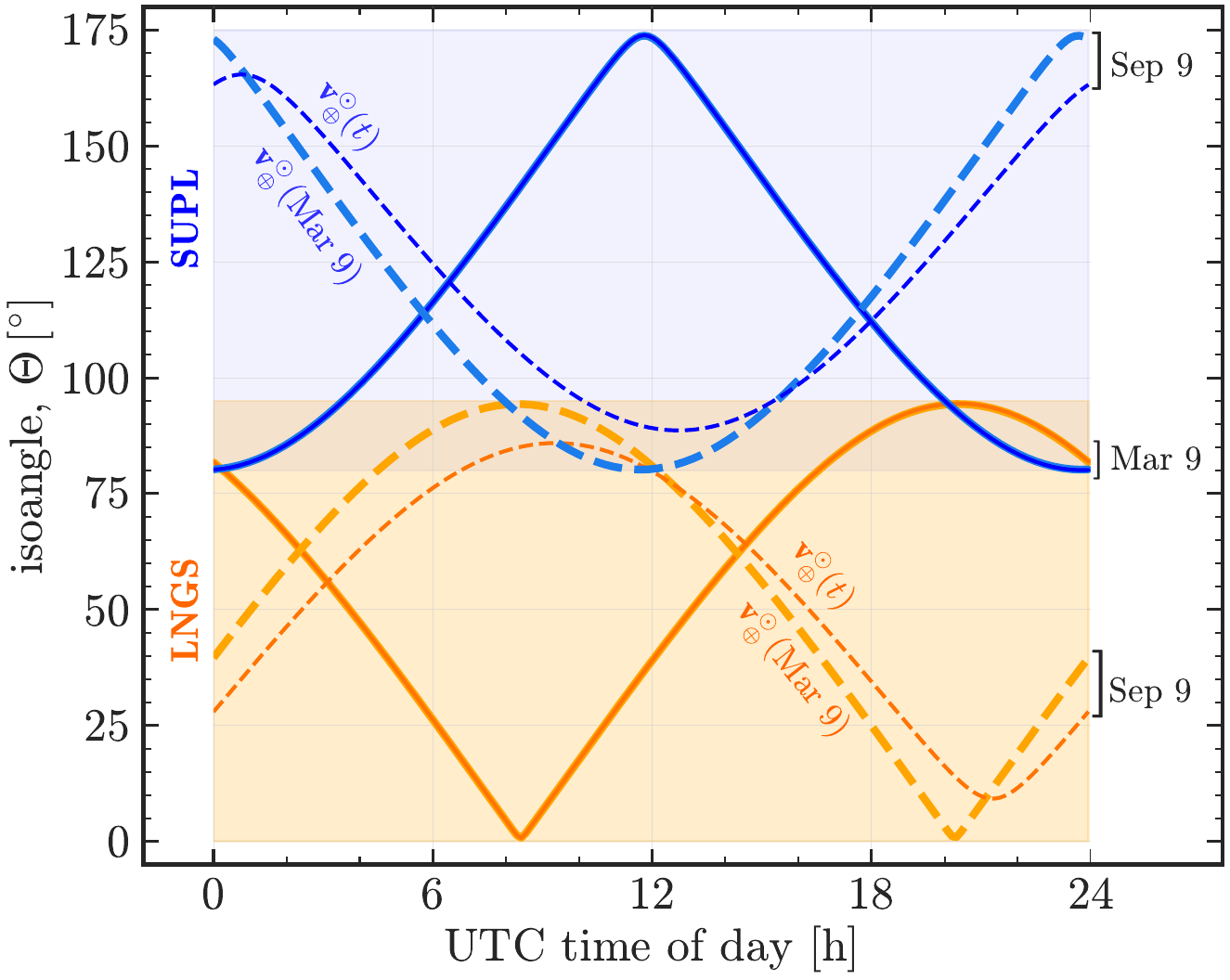}
\caption{Evolution of the isoangle $\Theta$ throughout the day at two detector locations (LNGS, shown in orange; and SUPL, shown in blue) and dates (March 9 and September 9, 2026). The thick lines use the fixed orbital-velocity vector evaluated on March 9 (see \Cref{eq:fixed_orbital_velocity}), which is adopted for the calculations in this work. The thin lines use the full season-dependent orbital velocity and are shown for comparison, see text for details.}
\label{fig:isoangle_evol}
\end{figure}

Figure~\ref{fig:isoangle_evol} shows the resulting evolution of $\Theta$ over one sidereal day at LNGS (Italy) and SUPL (Australia). The solid curves show March~9 using the March~9 orbital velocity. The thick dashed curves show September~9 while retaining the fixed March~9 orbital velocity adopted in our analysis, whereas the thin dashed curves show September~9 using the true season-dependent orbital velocity,
\begin{equation}
\mathbf{v}_{\oplus}^{\odot}(\text{September 9})
\simeq
(-29.0,0.1,-5.9)\ {\rm km\,s^{-1}}.
\label{eq:september_orbital_velocity}
\end{equation}
The solid and thick dashed curves therefore differ only through the date-dependent relation between UTC and sidereal phase, while the separation between the thick and thin dashed curves isolates the effect of the annual variation in $\mathbf{v}_{\oplus}^{\odot}(t)$.

For our analyses, we use the fixed March~9 velocity both to model the DM signal and to map background timestamps into isoangle space. The signal and background populations are therefore treated consistently, although the resulting $\Theta(t)$ does not capture the full seasonal dependence.

Two important effects can be further deduced from \Cref{fig:isoangle_evol}: 
\begin{itemize}
    \item Firstly, the range spanned by the modulating $\Theta(t)$ is as wide as $\sim$$90^{\circ}$. As previously seen from \Cref{fig:attenuated_velocities}, the Earth shielding effects depend strongly on $\Theta$. Combined  with the daily $\Theta$ evolution, this implies that the Galactic DM velocity distribution undergoes time-dependent attenuation, and the final $f(v$) incident on the detector varies periodically with time. While the hour of the day as such is \textit{not} the physical variable in which to interpret the degree of shielding, the $t \to \Theta(t)$ conversion can immediately predict the modified $f(v)$ for a given $(m_{\chi},\,\sigma_{\chi n
    }^{\mathrm{SI}})$.
    \item Secondly, laboratories in the Northern Hemisphere (such as SURF and LNGS) typically span $0^{\circ} \lesssim \Theta \lesssim 90^{\circ}$, while those in the Southern Hemisphere (such as SUPL) observe modulation in the  $90^{\circ} \lesssim \Theta \lesssim 180^{\circ}$ range. This stems from the fact that the Solar System motion throughout the Galaxy is approximately directed towards the Cygnus constellation \cite{Bandyopadhyay:2010zj}. The latter is visible in the Northern Hemisphere with the declination $\delta \approx 42^{\circ}$, and presents a fixed direction from which the DM wind appears to blow (up to the annual modulation effects). Therefore, to reach the detectors in the Southern Hemisphere, the Galactic DM particles need to traverse larger distances within the Earth compared to the detectors in the Northern Hemisphere. This results in a greater degree of Earth shielding of the DM flux in the South compared to the North.
\end{itemize}

\subsection{$\Theta$-Dependent Ionization Rates}\label{sec:scat_rate_calculation}

\begin{figure*}[t]
\includegraphics[width=\textwidth]{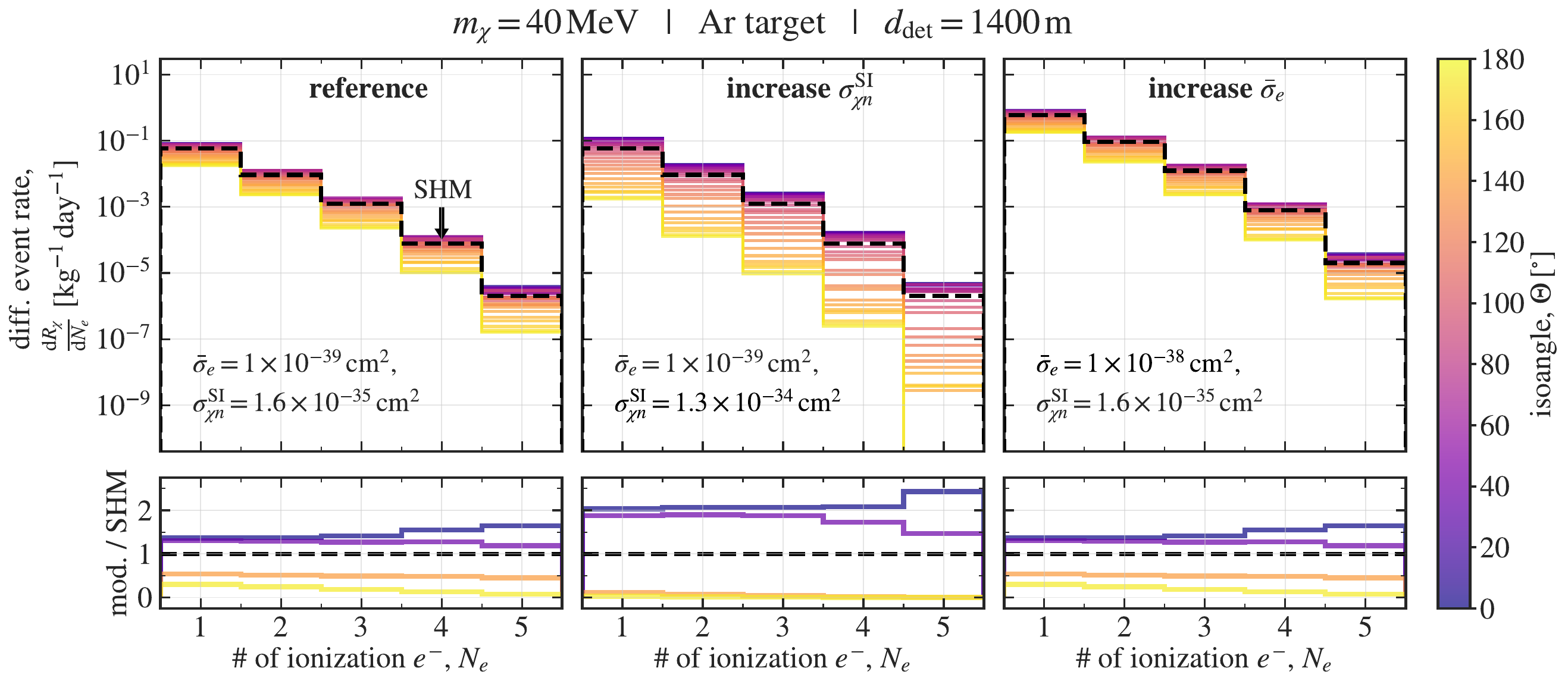}
\caption{DM-induced argon ionization spectra, $\frac{\mathrm{d}R_{\chi}}{\mathrm{d}N_e}$, shown as a function of the number of ionization electrons, $N_e$. The black dashed line corresponds to \Cref{eq:master_scattering_rate_eq} evaluated with the unattenuated (SHM) speed distribution $f(v)$, implying no DM scattering in the Earth. The colored lines are computed using the Earth-shielded $f(v\,|\,\Theta)$, which we extract from \texttt{DaMaSCUS} \cite{Emken:2017qmp,DaMaSCUS:2020xxx} in $6^{\circ}$-wide $\Theta$ bins. The three panels correspond to the different reference scenarios of $\sigma_{\chi n}^{\mathrm{SI}}$ and $\bar{\sigma}_e$, which control the scattering rates in the Earth and in the detector, respectively. The middle and the rightmost panels modify one quantity ($\sigma_{\chi n}^{\mathrm{SI}}$ or $\bar{\sigma}_e$) with respect to the leftmost panel, as indicated. The bottom panels show the ratio of the modulated rates to the SHM case for $\Theta$ bins centered at $3^{\circ}$, $39^{\circ}$, $141^{\circ}$, and $177^{\circ}$.}
\label{fig:modulated_dR_dNe_Ar}
\end{figure*}

\begin{figure*}[t]
\includegraphics[width=\textwidth]{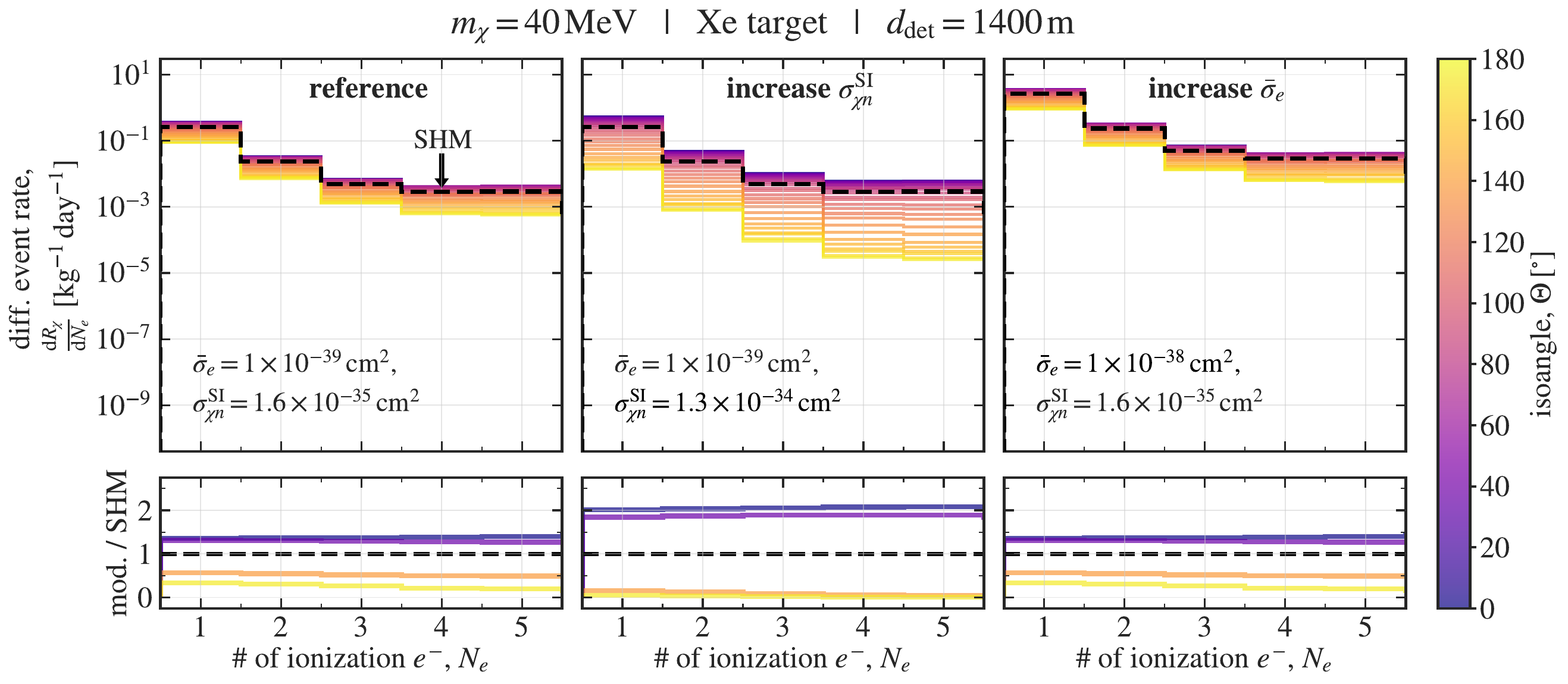}
\caption{Same as \Cref{fig:modulated_dR_dNe_Ar}, but for the xenon target.}
\label{fig:modulated_dR_dNe_Xe}
\end{figure*}

\begin{figure*}[t]
\includegraphics[width=0.8\textwidth]{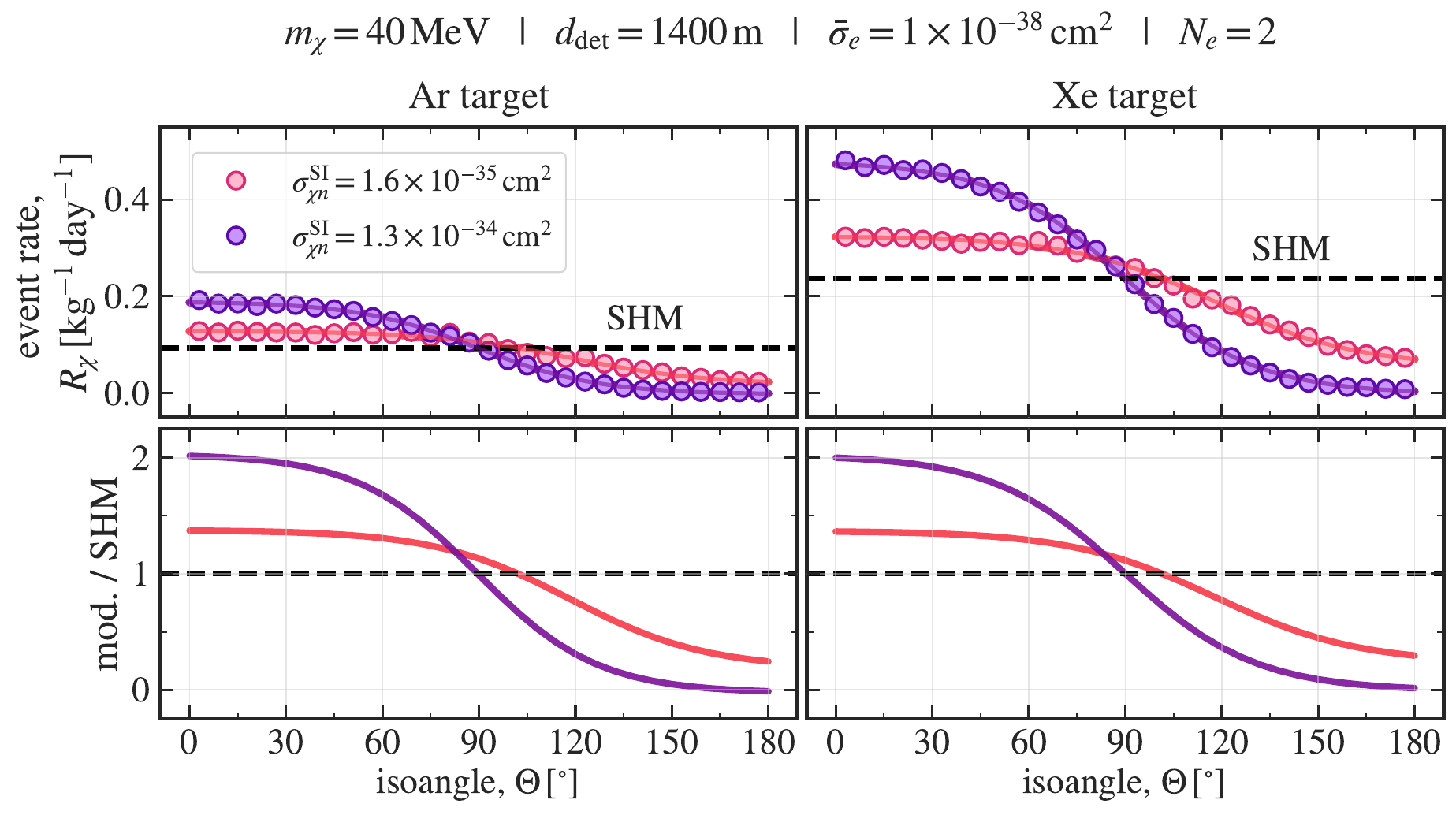}
\caption{Argon and xenon ionization rates at $N_e = 2$ shown as a function of the isoangle $\Theta$ for two reference values of the DM--nucleon scattering cross section, $\sigma_{\chi n}^{\mathrm{SI}}$. The scatter points are obtained by directly propagating the \texttt{DaMaSCUS} $\eta(v_{\mathrm{min}}\,|\,\Theta)$ into \Cref{eq:master_scattering_rate_eq}, while the solid lines show the fits with a hyperbolic tangent function. The bottom panels show the ratio of the fitted modulated rates to the SHM expectation (black dashed line).}
\label{fig:projected_mod_rates}
\end{figure*}

The dependence of the attenuated DM speed distributions $f(v\,|\Theta)$ on the isoangle propagates to the observable DM scattering rates inside the detector. We consider DM--electron scattering and subsequent atomic ionization as the detection channel. The corresponding differential rate per unit time and target mass is \cite{Essig:2011nj,DarkSide:2022knj}
\begin{align}
    \frac{\mathrm{d}R_{\chi}(\Theta)}{\mathrm{d}\ln E_{\mathrm{ER}}} &= \frac{\rho_{\chi}(\Theta)}{m_Nm_{\chi}} \frac{\bar{\sigma}_e}{8\mu_{\chi e}^2} \nonumber \times \\  &\sum_{n\ell}\,\int |f_{\mathrm{ion}}^{n\ell}(k', q)|^2\,|F_{\chi} (q)|^2\,\eta(v_{\mathrm{min}} | \Theta)\,q\,\mathrm{d}q,
\label{eq:master_scattering_rate_eq}
\end{align}
where
\begin{itemize}
    \item $E_{\mathrm{ER}}$ is the electron recoil energy;
    \item $m_N$ is the target nucleus mass;
    \item $\rho_{\chi}(\Theta)$ and $m_\chi$ are the local DM density (as a function of isoangle) and DM particle mass, respectively;
    \item $\bar{\sigma}_e$ is the reference DM--electron scattering cross section;
    \item $\mu_{\chi e} \equiv \frac{m_{\chi} m_e}{m_{\chi} + m_e}$ is the DM--electron reduced mass;
    \item  $|f_{\mathrm{ion}}^{n\ell}(k', q)|^2$ is the target ionization form factor, with $k' \equiv \sqrt{2m_e E_{{\mathrm{ER}}}}$ being the momentum of the outgoing electron, and $q$ the momentum transfer;
    \item $|F_{\chi}(q)|^2$ is the DM form factor defining the momentum transfer dependence of the interaction;
    \item $\eta(v_{\mathrm{min}} | \Theta) \equiv \int_{v_{\mathrm{min}}} \frac{f(v\,|\,\Theta)}{v}\,\mathrm{d}v$ is the average DM inverse speed above the minimum speed $v_{\mathrm{min}}$ required to induce an electron recoil with energy $E_{\mathrm{ER}}$. In our work, this quantity is obtained from the \texttt{DaMaSCUS} simulations (tabulated versus $v_{\mathrm{min}}$). Note that here the speed distributions $f(v\,|\,\Theta)$ are normalized to 1, such that the isoangle-dependent enhancement or depletion of the total DM flux is fully carried by the modified local DM density $\rho(\Theta)$.
\end{itemize}

The sum $\sum_{n\ell}$ in \Cref{eq:master_scattering_rate_eq} runs over the relevant electron shells $(n,\,\ell)$ for a given target atom, which are selected based on their binding energies $E_{\mathrm{bind}}^{n\ell}$. The latter have a direct impact on $v_{\mathrm{min}}$ \cite{Essig:2017kqs}, namely:
\begin{equation}
    v_{\mathrm{min}} = \frac{E_{\mathrm{ER}} +E_{\mathrm{bind}}^{n\ell}}{q} + \frac{q}{2m_{\chi}},
\end{equation}
meaning that large $E_{\mathrm{bind}}^{n\ell}$ raise the lower integration bound for $\eta(v_{\mathrm{min}})$ and reduce the resulting scattering rate. The following shells are ultimately taken into account for the two noble liquid targets studied in this work, with their binding energies given in parentheses: 
\begin{itemize}
    \item $2p$ (\SI{260.45}{eV}), $3s$ (\SI{34.76}{eV}), and $3p$ (\SI{16.08}{eV}) for Ar;
    \item $4s$ (\SI{213.77}{eV}), $4p$ (\SI{163.49}{eV}), $4d$ (\SI{75.59}{eV}), $5s$ (\SI{25.70}{eV}), and $5p$ (\SI{12.44}{eV}) for Xe.
\end{itemize}

In the above, we assume Roothaan-Hartree-Fock (RHF) orbitals as per Ref.~\cite{Bunge:1993jsz}. We note that in this non-relativistic framework, the sub-dominant relativistic effects such as the spin-orbit coupling splitting between the Argon orbitals $\{3p_{3/2},\,3p_{1/2}\}$ as well as the Xenon orbitals $\{5p_{3/2},\,5p_{1/2}\}$ and $\{4d_{5/2},\,4d_{3/2}\}$ are not taken into account. The RHF formalism is subsequently used to calculate the ionization form factors $|f_{\mathrm{ion}}^{n\ell}(k', q)|^2$, which we extract from the \texttt{DarkArt} package \cite{Catena:2019gfa,DarkArt:2022xxx}.

Figures~\ref{fig:modulated_dR_dNe_Ar} and~\ref{fig:modulated_dR_dNe_Xe} show examples of modulated differential ionization rates for argon and xenon targets, respectively. These are obtained by convolving ${\mathrm{d}R_{\chi}}\,/\,{\mathrm{d}\ln E_{\mathrm{ER}}}$ with the argon and xenon ionization yields following \cite{DarkSide-50:2022qzh,DarkSide:2022knj,Essig:2017kqs,Catena:2025ung}. The ensuing quantity is the differential ionization spectrum ${\mathrm{d}R}_{\chi}\,/\,{\mathrm{d}N_e}$, where $N_e$ is the number of ionization electrons. Note that the argon ionization yield model from Refs.~\cite{DarkSide-50:2022qzh,DarkSide:2022knj} returns $N_e$ as a continuous quantity, and our interpretation of $N_e = N_e'$ is an integral over the $[N_e',\,N_e'+1]$ range. The xenon ionization yield model from Refs.~\cite{Essig:2017kqs,Catena:2025ung} returns a probability mass function of integer $N_e$.

Importantly, \Cref{fig:modulated_dR_dNe_Ar,fig:modulated_dR_dNe_Xe} highlight that DM--electron and DM--nucleon cross sections have distinctly different effects on the observable spectra. While $\bar{\sigma}_e$ simply rescales the overall event rate in the detector, $\sigma_{\chi n}^{\mathrm{SI}}$ controls the degree of Earth shielding and therefore impacts the spread of the $\Theta$-dependent event rates about the SHM expectation.  

Figure~\ref{fig:projected_mod_rates} extracts the ionization rates at $N_e = 2$ from \Cref{fig:modulated_dR_dNe_Ar,fig:modulated_dR_dNe_Xe} and projects them onto the $\Theta$ axis. This projection further emphasizes that the higher value of $\sigma_{\chi n}^{\mathrm{SI}}$ results in a stronger $\Theta$ dependence and thus a more prominent modulation signature compared to the lower value. In the figure, we show both scatter points, as obtained from \texttt{DaMaSCUS}, as well as fits for the $\Theta$-dependent ionization rates with a four-parameter hyperbolic tangent function (following Ref.~\cite{Bertou:2025adb}). This is done to smooth the rates obtained from the Monte Carlo-derived DM speed distributions and to ultimately build rate interpolators on the $(m_{\chi},\,\mathrm{\sigma}_{\chi n}^{\mathrm{SI}})$ grids. In the few cases of limited simulation statistics or very weak modulation signatures, we adopt a parabolic parameterization to improve fit convergence.

\section{Statistical Analysis}
\label{sec:stat_analysis}

\begin{figure}[t]
\includegraphics[width=\columnwidth]{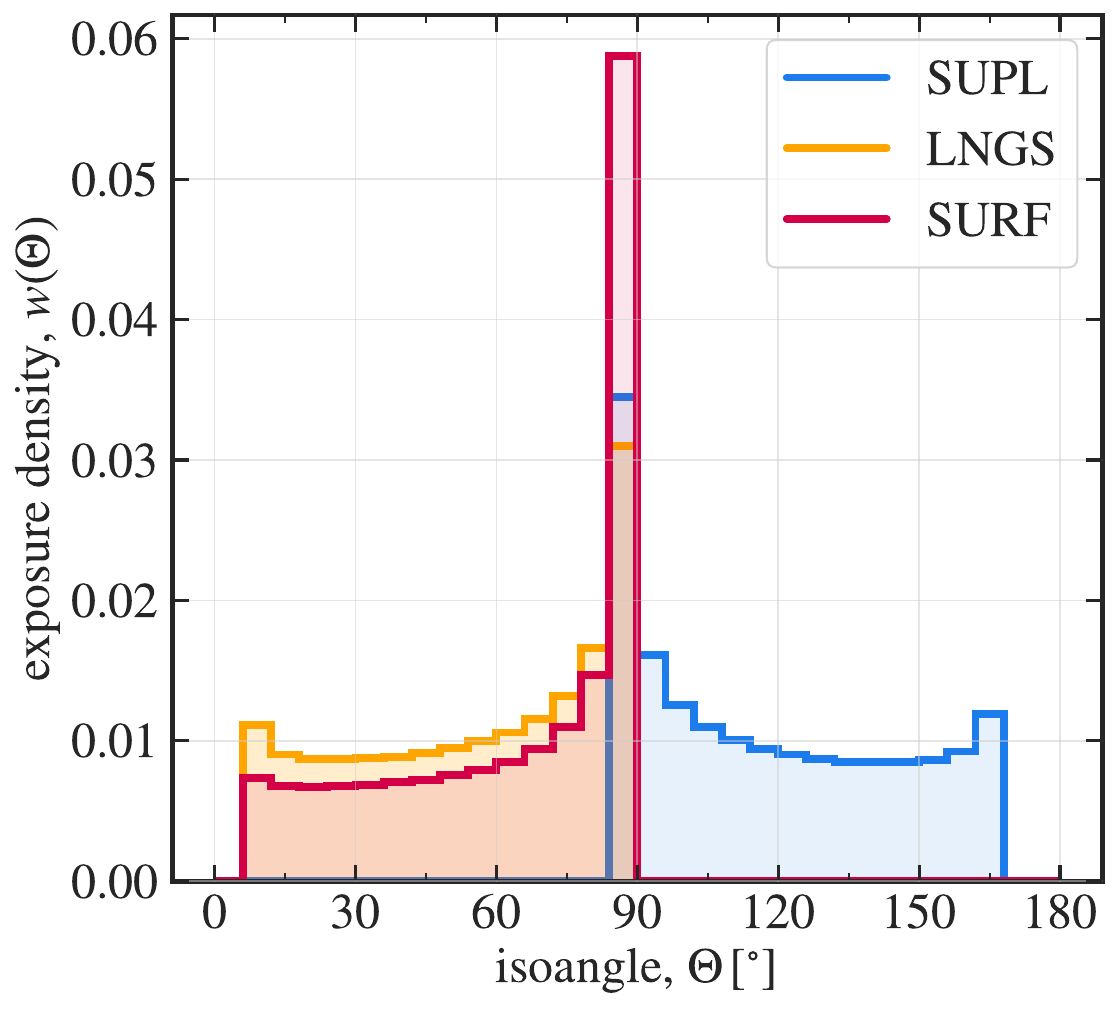}
\caption{Daily isoangle exposure densities $w(\Theta)$ at three laboratory locations: SUPL, LNGS, and SURF. The distributions are obtained by uniformly sampling time over one sidereal day and mapping each timestamp to $\Theta$ using \Cref{eq:isoangle_definition}; see also \Cref{fig:isoangle_evol}.}
\label{fig:isoangle_exposure}
\end{figure}

\begin{figure}[t]
\includegraphics[width=\columnwidth]{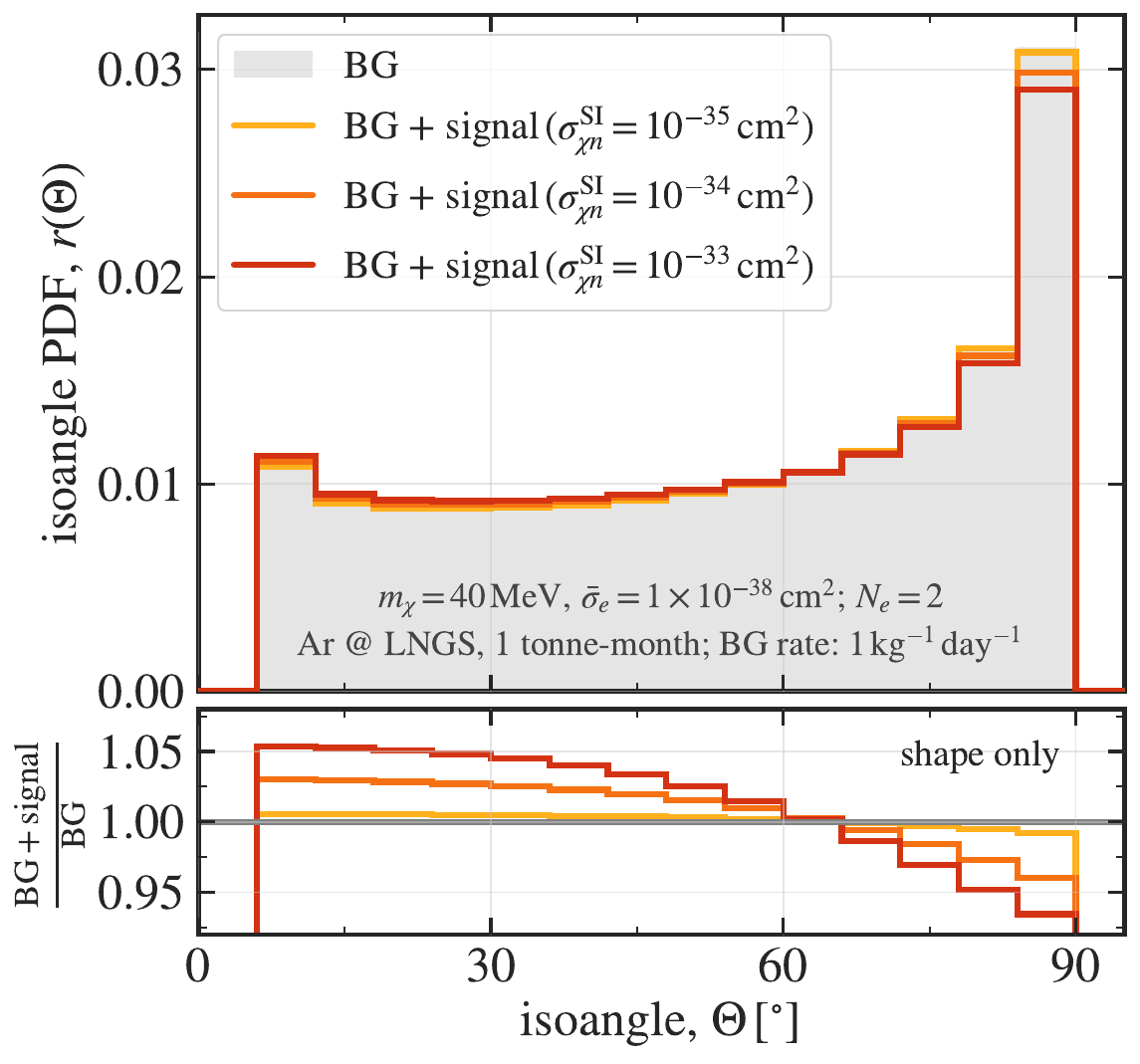}
\caption{Isoangle probability densities $r(\Theta)$ at LNGS for the background-only hypothesis, shown by the filled grey histogram, and three signal-plus-background hypotheses, shown by the outlined histograms. Increasing $\sigma_{\chi n}^{\mathrm{SI}}$ strengthens the Earth-shielding effect and produces a larger departure from the background-only shape.}
\label{fig:modified_theta_hists_examples}
\end{figure}

The rates derived in \Cref{sec:scat_rate_calculation} describe the DM signal as a function of isoangle $\Theta$, but do not account for how frequently a detector samples each value of $\Theta$ over a sidereal day. We first incorporate this location-dependent exposure (which was not included in previous work), then define a statistic that quantifies the resulting $\Theta$-distribution shape, and finally calibrate this statistic using Poisson realizations of the expected event distributions to construct exclusion limits.

\subsection{Location-Dependent Isoangle Exposure}
\label{sec:isoangle_exposure_statistics}

Because $\Theta(t)$ varies nonuniformly throughout the day, a detector does not receive equal exposure to all accessible isoangles. We characterize this effect by the exposure density $w(\Theta)$, normalized such that
\begin{equation}
\int w(\Theta)\,\mathrm{d}\Theta
\simeq
\sum_i w(\Theta_i)\Delta\Theta
=1,
\end{equation}
where $\Theta_i$ denotes the isoangle-bin center. 

Figure~\ref{fig:isoangle_exposure} shows $w(\Theta)$ at SUPL, LNGS, and SURF, obtained by binning the values of $\Theta(t)$ sampled uniformly over one sidereal day. We use a bin width $\Delta\Theta=6^\circ$, matching the isodetection rings simulated with \texttt{DaMaSCUS}.

Recorded event timestamps can be mapped into isoangle space using \Cref{eq:isoangle_definition}. A time-independent background therefore acquires a $\Theta$ distribution determined solely by $w(\Theta)$, whereas the DM signal contains both this exposure dependence and the intrinsic Earth-shielding dependence of $R_\chi(N_e,\Theta)$. For a $\Theta$-independent background rate $R_{\mathrm{bg}}(N_e)$, the exposure-weighted background and signal rate densities are
\begin{equation}
\widetilde{R}_{\mathrm{bg}}(N_e,\Theta_i)
=
R_{\mathrm{bg}}(N_e)\,w(\Theta_i),
\label{eq:exposure_aware_rate_background}
\end{equation}
and
\begin{equation}
\widetilde{R}_{\chi}(N_e,\Theta_i)
=
R_{\chi}(N_e,\Theta_i)\,w(\Theta_i),
\label{eq:exposure_aware_rate_signal}
\end{equation}
respectively. The expected rate in bin $\Theta_i$ is obtained by multiplying these quantities by $\Delta\Theta$. The total exposure-weighted rate density is
\begin{equation}
\widetilde{R}_{\mathrm{tot}}(N_e,\Theta_i)
=
\widetilde{R}_{\mathrm{bg}}(N_e,\Theta_i)
+
\widetilde{R}_{\chi}(N_e,\Theta_i).
\end{equation}

For a fixed $N_e$ bin, we define the corresponding normalized isoangle probability density as
\begin{equation}
r(\Theta_i)
\equiv
\frac{\widetilde{R}(\Theta_i)}
{\displaystyle\sum_j
\widetilde{R}(\Theta_j)\Delta\Theta}.
\label{eq:isoangle_pdf}
\end{equation}
The difference between the background-only and signal-plus-background distributions depends on the relative signal and background rates. The DM--nucleon cross section $\sigma_{\chi n}^{\mathrm{SI}}$ controls the intrinsic $\Theta$ dependence produced by Earth shielding, while the DM--electron cross section $\bar{\sigma}_e$ controls the overall signal normalization.

Figure~\ref{fig:modified_theta_hists_examples} compares the background-only and signal-plus-background $\Theta$ distributions at LNGS for several values of $\sigma_{\chi n}^{\mathrm{SI}}$ at a fixed $\bar{\sigma}_e$. The characteristic enhancement at small $\Theta$ and suppression at large $\Theta$ remain visible after including both the background contribution and the location-dependent exposure.

\subsection{Isoangle Shape Statistic}
\label{sec:shape_statistic}

We next define a statistic that quantifies the difference between the background-only and signal-plus-background isoangle shapes. For a binned probability density function (PDF) $r(\Theta_i)$, the corresponding cumulative distribution function (CDF) is
\begin{equation}
\xi(\Theta_i)
=
\sum_{j\leq i}
r(\Theta_j)\Delta\Theta.
\label{eq:isoangle_cdf}
\end{equation}
The conventional CDF comparison tool is the Kolmogorov-Smirnov (KS) statistic \cite{Kolmogorov:1933xxx,Smirnov:1933xxx}, which evaluates the maximum absolute separation between two cumulative distribution functions. For the background-only and signal-plus-background CDFs, $\xi_{\mathrm{b}}$ and $\xi_{\mathrm{sb}}$, this is
\begin{equation}
D_{\mathrm{KS}}^{\mathrm{max}}
=
\underset{i}{\mathrm{sup}}
\left|
\xi_{\mathrm{sb}}(\Theta_i)
-
\xi_{\mathrm{b}}(\Theta_i)
\right|.
\label{eq:ks_test_statistic_maximum}
\end{equation}
Instead, we use the integrated absolute CDF difference,
\begin{equation}
D_{\mathrm{isoang}}
=
\sum_i
\left|
\xi_{\mathrm{sb}}(\Theta_i)
-
\xi_{\mathrm{b}}(\Theta_i)
\right|
\Delta\Theta,
\label{eq:ks_test_statistic_integral}
\end{equation}
which is equivalent to the one-dimensional Wasserstein-1 distance~\cite{Aaditya:2017xxx}. Unlike the conventional KS statistic, which depends only on the largest local CDF separation, $D_{\mathrm{isoang}}$ accumulates differences over the full isoangle range and is therefore less sensitive to isolated statistical fluctuations.

\begin{figure}[t]
\includegraphics[width=\columnwidth]{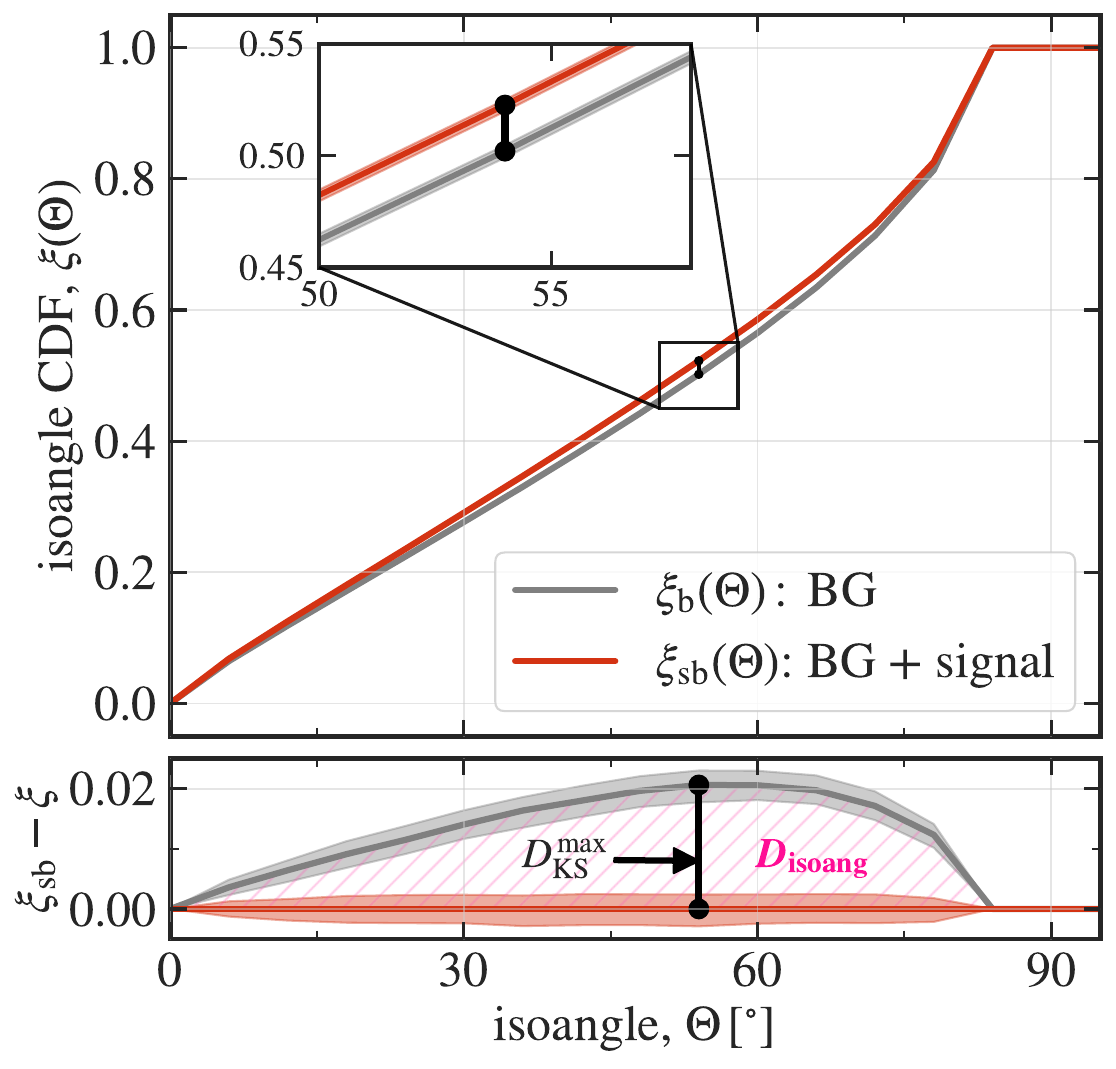}
\caption{Comparison of the background-only and signal-plus-background isoangle CDFs, $\xi_{\mathrm{b}}(\Theta)$ and $\xi_{\mathrm{sb}}(\Theta)$. The conventional KS statistic $D_{\mathrm{KS}}^{\mathrm{max}}$ is the maximum vertical separation between the CDFs, while the isoangle shape statistic $D_{\mathrm{isoang}}$ used in this work is the integrated absolute separation. The solid curves show the Asimov CDFs, and the shaded regions indicate the intervals between the $16\%$ and $84\%$ percentiles of their fluctuated realizations.}
\label{fig:ks_test_explanation}
\end{figure}

Figure~\ref{fig:ks_test_explanation} illustrates the background-only and signal-plus-background CDFs and compares the conventional maximum separation with the integrated distance adopted here. 

\subsection{Hypothesis Testing and Exclusion Procedure}
\label{sec:hypothesis_testing}

We consider the background-only hypothesis, $H_{\rm b}$, and a specified signal-plus-background hypothesis, $H_{\rm sb}$. For each point in the signal parameter space $(m_\chi,\sigma_{\chi n}^{\rm SI},\bar{\sigma}_e)$, we construct two distributions of $D_{\rm isoang}$ using Poisson pseudoexperiments.

Under the background-only hypothesis, we generate fluctuated templates
$\widetilde{R}_{\mathrm{bg}}^{*}(\Theta)$
and calculate their isoangle shape distance from the fixed Asimov signal-plus-background prediction
$\widetilde{R}_{\mathrm{tot}}(\Theta)$.
The resulting test-statistic distribution is denoted by
\begin{equation}
\mathcal{K}_{\mathrm{b}}(D_{\mathrm{isoang}})
\equiv
\mathrm{PDF}\left[
D_{\mathrm{isoang}}
\left(
H_{\mathrm{sb}},H_{\mathrm{b}}^{*}
\right)
\right].
\end{equation}
As the signal strength or Earth-shielding effect increases, the background-only realizations become more separated from the signal-plus-background prediction, shifting $\mathcal{K}_{\mathrm{b}}$ toward larger values of $D_{\mathrm{isoang}}$.

Similarly, under the signal-plus-background hypothesis, we generate fluctuated templates
$\widetilde{R}_{\mathrm{tot}}^{*}(\Theta)$
and compare them with the corresponding Asimov prediction. This gives
\begin{equation}
\mathcal{K}_{\mathrm{sb}}(D_{\mathrm{isoang}})
\equiv
\mathrm{PDF}\left[
D_{\mathrm{isoang}}
\left(
H_{\mathrm{sb}},H_{\mathrm{sb}}^{*}
\right)
\right],
\end{equation}
whose width reflects the Poisson fluctuations expected under the alternative hypothesis. The separation and overlap of
$\mathcal{K}_{\mathrm{b}}$
and
$\mathcal{K}_{\mathrm{sb}}$
determine the discriminating power of the isoangle shape test.

\begin{figure*}[t]
\includegraphics[width=\textwidth]{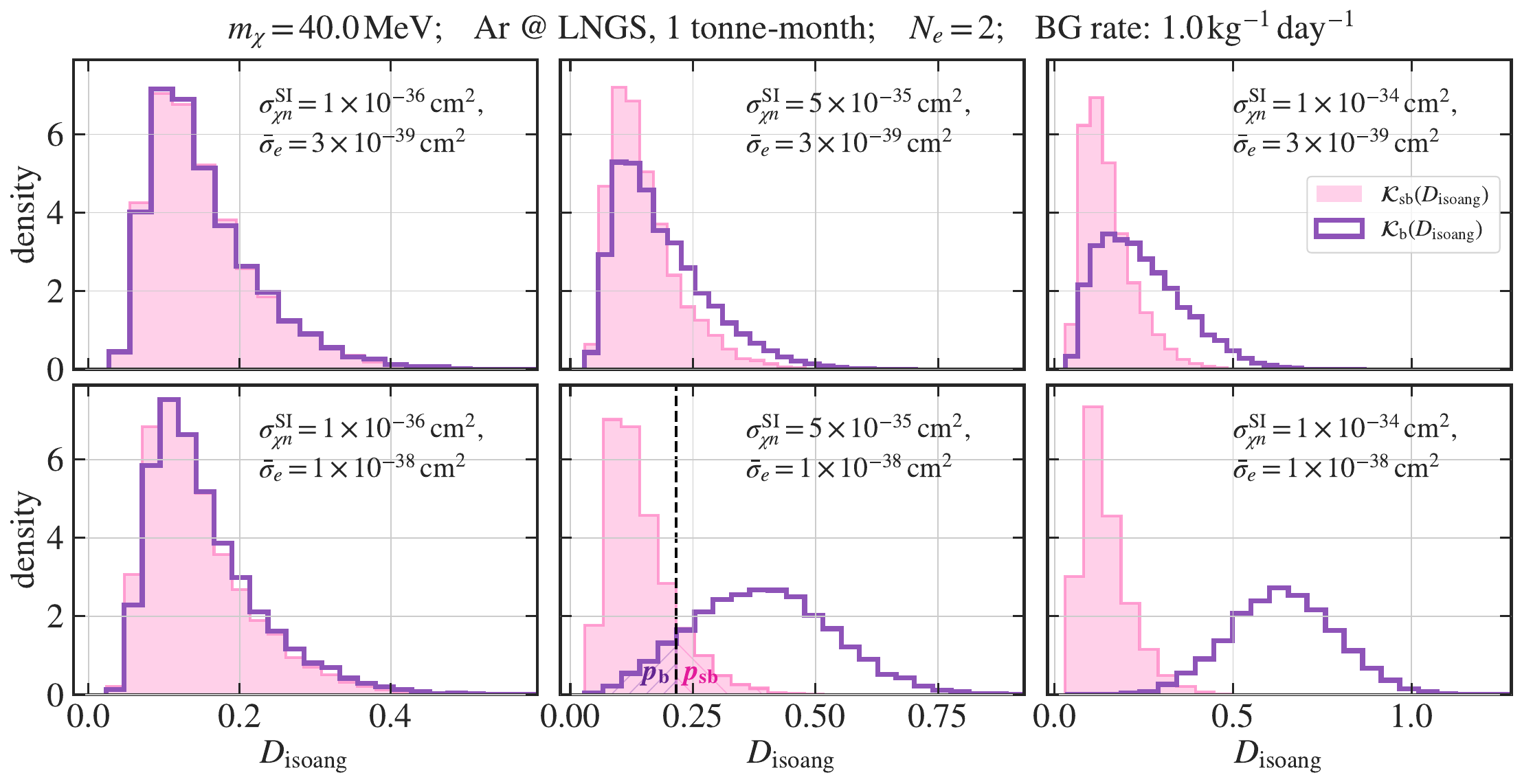}
\caption{Distributions of the isoangle shape statistic $D_{\mathrm{isoang}}$ for six signal hypotheses at fixed DM mass and detector configuration. The outlined histograms show $\mathcal{K}_{\mathrm{b}}$, obtained by comparing the Asimov signal-plus-background prediction with Poisson-fluctuated background-only realizations. The filled histograms show $\mathcal{K}_{\mathrm{sb}}$, obtained from Poisson-fluctuated signal-plus-background realizations. The bottom-center panel illustrates the construction of the CL$_s$ shape-test $p$-value  (see text for details).}
\label{fig:ks_distances_example}
\end{figure*}

For observed data, we define
$D_{\mathrm{isoang}}^{\mathrm{obs}}$
as the distance between the observed isoangle CDF and that predicted by the signal-plus-background model. The probability of obtaining a distance at least as large under the alternative hypothesis is
\begin{equation}
p_{\mathrm{sb}}
=
\int_{D_{\mathrm{isoang}}^{\mathrm{obs}}}^{\infty}
\mathcal{K}_{\mathrm{sb}}(D_{\mathrm{isoang}})
\,\mathrm{d}D_{\mathrm{isoang}}.
\label{eq:pvalue_psb}
\end{equation}
This quantity measures the compatibility of the observed data with the specified signal-plus-background model. A large value of
$D_{\mathrm{isoang}}^{\mathrm{obs}}$
relative to
$\mathcal{K}_{\mathrm{sb}}$
indicates that the model is disfavored.

We also define
\begin{equation}
p_{\mathrm{b}}
=
\int_0^{D_{\mathrm{isoang}}^{\mathrm{obs}}}
\mathcal{K}_{\mathrm{b}}(D_{\mathrm{isoang}})
\,\mathrm{d}D_{\mathrm{isoang}},
\label{eq:cls_pb}
\end{equation}
so that $1-p_{\mathrm{b}}$ is the probability, under the background-only hypothesis, of obtaining a distance at least as large as the observed value. The CL$_\mathrm{s}$ $p$-value resulting from the isoangle shape test is then \cite{Cowan:1998ji}
\begin{equation}
p_{\mathrm{isoang}}
=
\frac{p_{\mathrm{sb}}}{1-p_{\mathrm{b}}}.
\label{eq:cls_pvalue}
\end{equation}

We evaluate $p_{\mathrm{isoang}}$ throughout the
$(m_\chi,\sigma_{\chi n}^{\mathrm{SI}},\bar{\sigma}_e)$
parameter grid. For fixed $\bar{\sigma}_e$, the resulting $90\%$ confidence-level exclusion contour in the
$(m_\chi,\sigma_{\chi n}^{\mathrm{SI}})$
plane identifies signal models whose predicted isoangle distributions are incompatible with the observed data. In the absence of a daily modulation signal, parameter points satisfying
\begin{equation}
p_{\mathrm{isoang}}<0.1
\end{equation}
are excluded at $90\%$ C.L.

Figure~\ref{fig:ks_distances_example} shows the distributions $\mathcal{K}_{\mathrm{b}}$ and $\mathcal{K}_{\mathrm{sb}}$ for representative signal hypotheses. As the DM--electron and DM--nucleon cross sections increase, the separation between the two distributions generally grows, strengthening the discriminating power of the shape test. The bottom-center panel illustrates the construction of the CL$_\mathrm{s}$ $p$-value described above.

\section{Sensitivity Projections for Liquid Nobles}
\label{sec:results_general_nobles}

\subsection{Projection Assumptions}
\label{sec:projection_assumptions}

We consider two scenarios for the DM--electron and DM--nucleon interactions, as detailed in Appendix~\ref{app:models}. In the first, both interactions are mediated by a heavy dark photon, and the corresponding cross sections are related through \Cref{eq:dark_photon_xsec_relation}. In the second, the electron and nucleon interactions arise from distinct mediators, so that $\bar{\sigma}_e$ and $\sigma_{\chi n}^{\mathrm{SI}}$ can vary independently.

\begin{figure*}[t]
\includegraphics[width=\textwidth]{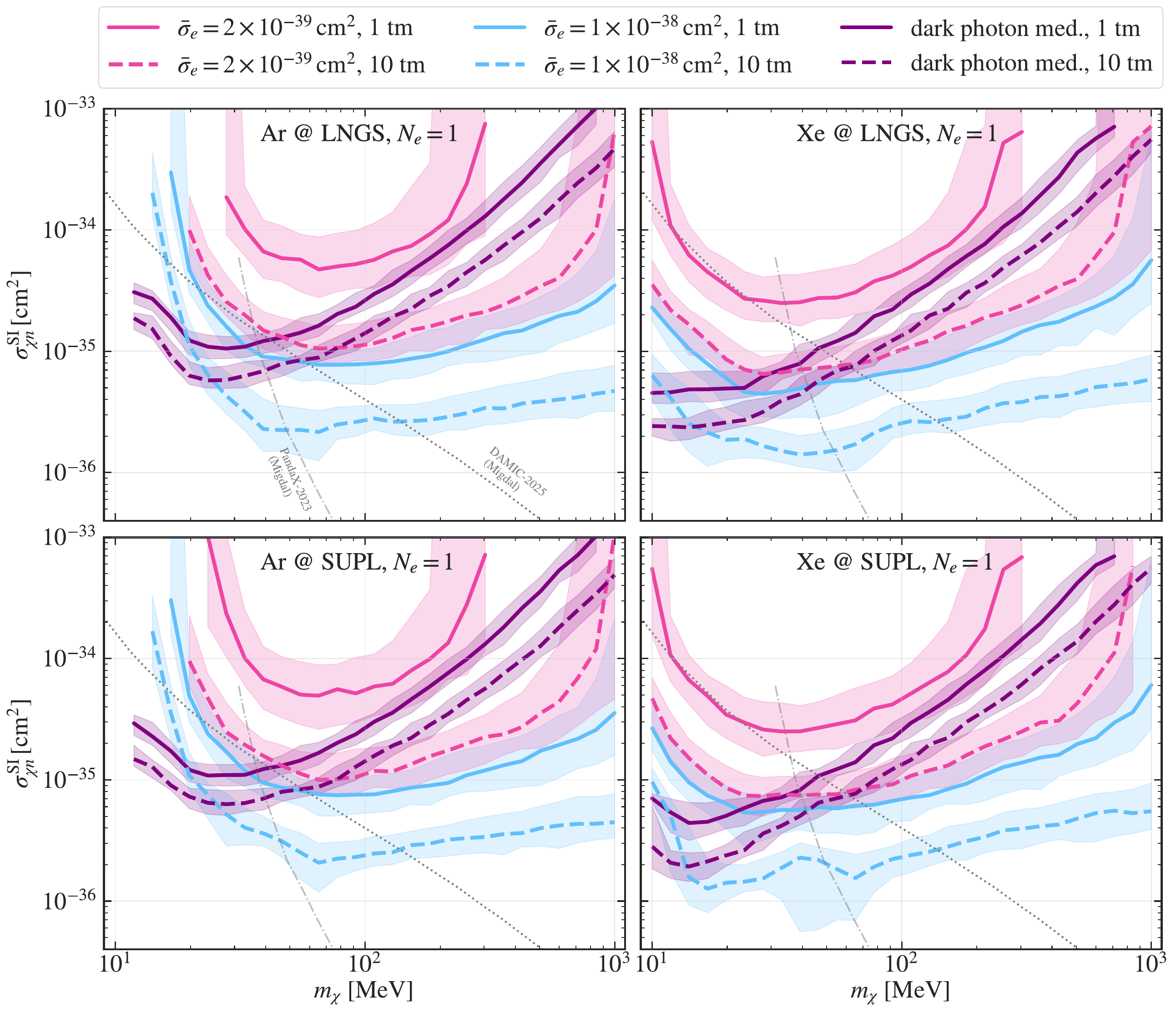}
\caption{Projected $90\%$ C.L. exclusion limits on $\sigma_{\chi n}^{\mathrm{SI}}$ using the $N_e=1$ ionization bin for argon and xenon targets at LNGS and SUPL. For each target and laboratory, contours are shown for exposures of \SI{1}{tm} and \SI{10}{tm} and for three DM--electron interaction benchmarks: two fixed values of $\bar{\sigma}_e$ and the dark-photon model relating $\bar{\sigma}_e$ to $\sigma_{\chi n}^{\mathrm{SI}}$ through \Cref{eq:dark_photon_xsec_relation}. The thick curves indicate where the median CL$_s$ shape-test $p$-value reaches 0.1. The shaded bands show the approximate $\pm1\sigma$ statistical variation, bounded by the cross sections at which the 16th and 84th percentiles of the $p$-value distribution reach 0.1.}
\label{fig:projections_Ne_1}
\end{figure*}

The version of \texttt{DaMaSCUS} used here parameterizes the coherent DM--nucleus cross section as
$\sigma_{\chi N}\propto\sigma_{\chi p}Z^2$,
where $\sigma_{\chi p}$ is the DM--proton cross section and $Z$ is the nuclear charge. It further assumes $Z/A=0.5$ for all terrestrial nuclei. To express the results in terms of the conventional spin-independent per-nucleon cross section, defined through
$\sigma_{\chi N}\propto\sigma_{\chi n}^{\mathrm{SI}}A^2$,
we therefore identify
\begin{equation}
\sigma_{\chi n}^{\mathrm{SI}}
\simeq
\sigma_{\chi p}
\left(\frac{Z}{A}\right)^2
=
\frac{\sigma_{\chi p}}{4}.
\label{eq:damascus_cross_section_conversion}
\end{equation}

For the generic argon and xenon projections, we consider detector locations at LNGS and SUPL and exposures of \SI{1}{tm} (tonne-month) and \SI{10}{tm}. We analyze separately the lowest two ionization bins, $N_e=1$ and $N_e=2$. For xenon, we adopt the background rates of Ref.~\cite{Bertou:2025adb}. For argon, we use the spurious-electron background extrapolated from the DarkSide-50 data to $N_e=1$ using \Cref{eq:se_model}; see Appendix~\ref{app:se_background_modelling}. The assumed rates are summarized in \Cref{tab:bg_rates}.

\begin{table}[]
\renewcommand{\arraystretch}{1.3}
\setlength{\tabcolsep}{10pt}
\begin{tabular}{lll}
\# of ionization $e^{-}$ & Ar                                        & Xe \\[1pt]  \cline{1-3} 
$N_e = 1$                & 5.1  & 3.0  \\
$N_e = 2$                & 0.4 & 0.1 \\ \cline{1-3}\\
\end{tabular}
\caption{Argon and xenon background rates (in $\mathrm{kg}^{-1}\,\mathrm{day}^{-1}$) assumed for sensitivity projections.\label{tab:bg_rates}}
\end{table}

\begin{figure*}[t]
\includegraphics[width=\textwidth]{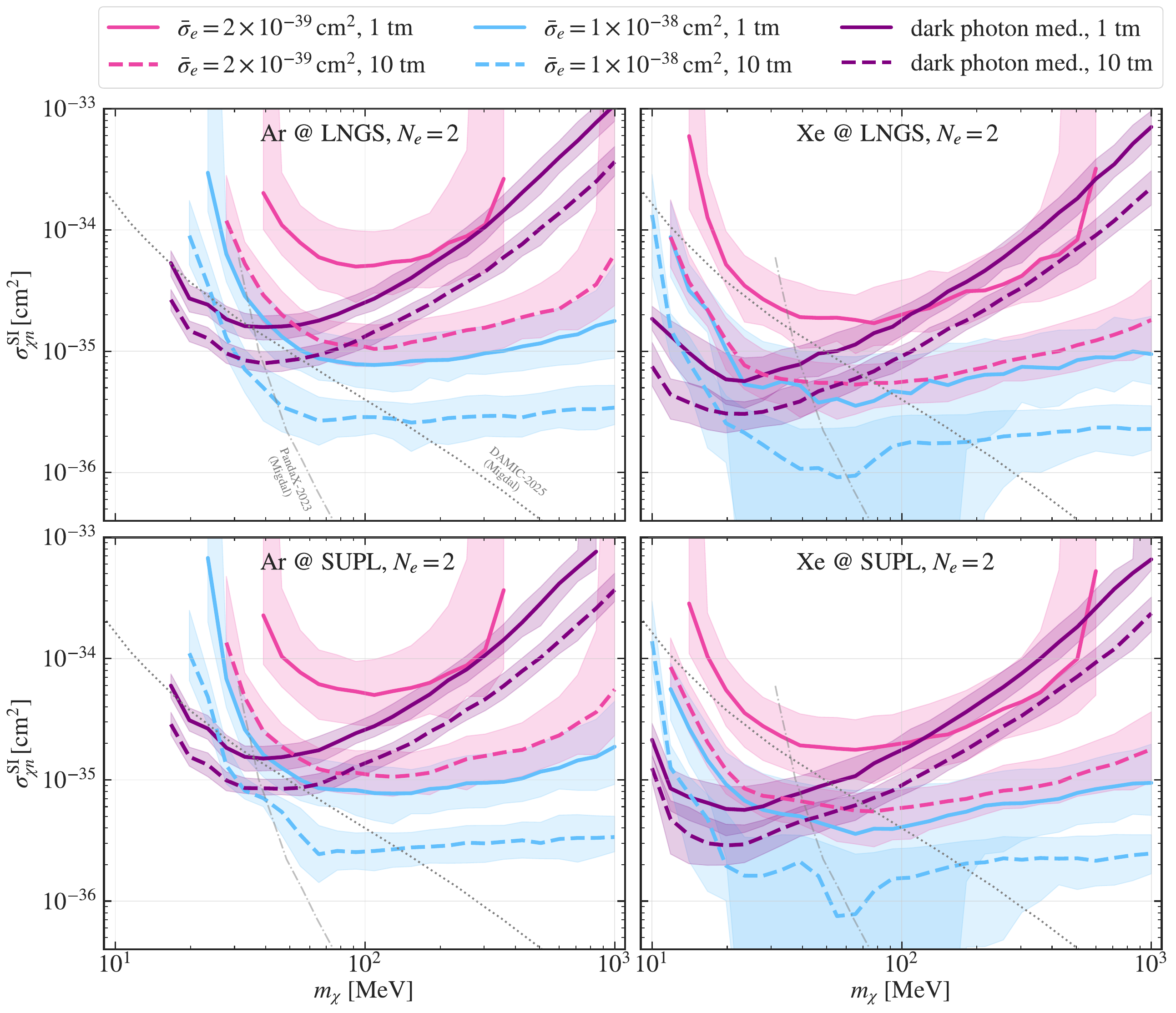}
\caption{Same as \Cref{fig:projections_Ne_1}, but using the $N_e=2$ ionization bin.}
\label{fig:projections_Ne_2}
\end{figure*}

\begin{figure*}[t]
\includegraphics[width=\textwidth]{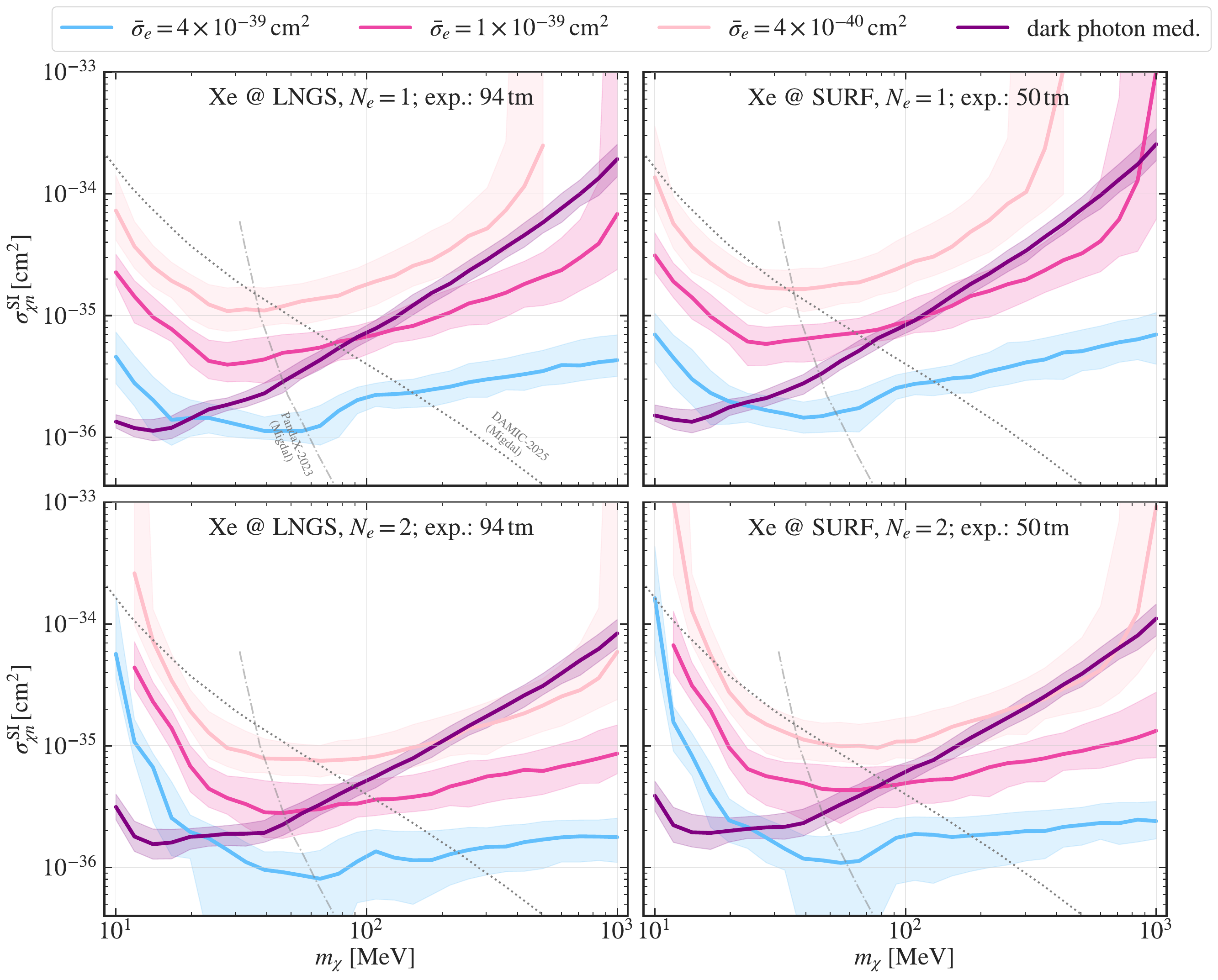}
\caption{Projected $90\%$ C.L. exclusion limits on
$\sigma_{\chi n}^{\mathrm{SI}}$
for a XENONnT-like detector at LNGS and an LZ-like detector at SURF. The exposures correspond to those reported by XENONnT~\cite{XENON:2026qow} and LZ~\cite{LZ:2024zvo}, respectively. The xenon background rates in \Cref{tab:bg_rates} are assumed for both detector configurations. The interaction benchmarks and statistical bands follow the conventions of \Cref{fig:projections_Ne_1}.}
\label{fig:projections_nT_LZ}
\end{figure*}

\subsection{Generic Argon and Xenon Projections}
\label{sec:generic_noble_projections}

Figures~\ref{fig:projections_Ne_1} and \ref{fig:projections_Ne_2} show the projected $90\%$ C.L. exclusions on intraterrestrial DM--nucleon scattering inferred through DM--electron ionization signals. The two figures correspond to the $N_e=1$ and $N_e=2$ bins, respectively. For each target and laboratory, we show projections for two exposures and for both the dark-photon scenario and representative fixed values of $\bar{\sigma}_e$ in the decoupled-interaction scenario. For Fig.~\ref{fig:projections_Ne_2} which shows the $N_e=2$ results, note that the reduced background for the xenon target in this $N_e$ bin implies that the DM signal constitutes a substantial fraction of the total event rate at large DM–electron cross sections. Consequently, for the $\bar{\sigma}_e = 10^{-38}\,\mathrm{cm}^2$ case shown here, the 16$^{\mathrm{th}}$ percentile of the $p$-value distribution never reaches 10\%, producing the characteristic ``dip'' in the lower boundary of the shaded band.

For comparison with existing searches, we overlay constraints on
$\sigma_{\chi n}^{\mathrm{SI}}$
from PandaX~\cite{PandaX:2023xgl} and DAMIC-M~\cite{DAMIC-M:2025luv}, obtained using the Migdal effect~\cite{Migdal:1941xxx}. The daily-modulation projections are competitive over approximately
$20$--\SI{40}{MeV}
for argon and
$10$--\SI{60}{MeV}
for xenon. Depending on the exposure and the assumed value of $\bar{\sigma}_e$, the projected sensitivity can improve upon the DAMIC-M limits by up to approximately one order of magnitude.

The relative performance of the $N_e=1$ and $N_e=2$ analyses is governed by the competition between signal acceptance and background. The lower threshold of the $N_e=1$ bin provides access to lighter DM and generally increases the signal rate, whereas the lower background in the $N_e=2$ bin can improve sensitivity where sufficient signal remains above threshold.

\subsection{Experiment-Scale Projections}
\label{sec:experiment_scale_projections}

The exposures adopted above are readily accessible to current and upcoming liquid-noble experiments. DarkSide-50 accumulated approximately \SI{0.42}{tm} liquid-argon exposure over 653 days \cite{DarkSide:2022knj,DarkSide-50:2022qzh,DarkSide:2022dhx}; we present a dedicated DarkSide-50 case study in \Cref{sec:darkside_case}. Its successor, DarkSide-20k, will operate with approximately \SI{20}{t} of liquid argon~\cite{DarkSide-20k:2024yfq}, reaching exposures of \SI{1}{tm} and \SI{10}{tm} in approximately 1.5 and 15 days, respectively.

Existing xenon experiments have already accumulated substantially larger exposures. XENONnT, with a \SI{5.9}{t} liquid-xenon target at LNGS, has collected \SI{7.83}{ty}, or approximately \SI{94}{tm}, of data~\cite{XENON:2026qow}. LZ at SURF has reported results based on \SI{4.2}{ty}, corresponding to approximately \SI{50.4}{tm}~\cite{LZ:2024zvo}. These exposures exceed the largest generic benchmark considered above by factors of approximately nine and five, respectively.

Figure~\ref{fig:projections_nT_LZ} shows projections for XENONnT-like and LZ-like detectors using their reported exposures. We adopt the xenon background rates in \Cref{tab:bg_rates} for both detector configurations.

When the DM--electron and DM--nucleon interactions are independent, the expected signal count scales with the product of the exposure and $\bar{\sigma}_e$. Increasing either quantity by a factor $k$ therefore increases the expected number of signal events by the same factor. Their effects on the sensitivity are nevertheless not identical: increasing the exposure also increases the expected background count, whereas increasing $\bar{\sigma}_e$ changes only the signal. Consequently, at a fixed background rate per unit exposure, increasing $\bar{\sigma}_e$ by a factor $k$ generally produces a stronger improvement in the limit on $\sigma_{\chi n}^{\mathrm{SI}}$ than increasing the exposure by the same factor. This distinction is important when comparing the experiment-scale projections in \Cref{fig:projections_nT_LZ} with the generic benchmarks in \Cref{fig:projections_Ne_1,fig:projections_Ne_2}.

\section{Case study: Limit-setting for Darkside-50}\label{sec:darkside_case}

We demonstrate the application of the statistical framework developed in \Cref{sec:stat_analysis} using the 653-day DarkSide-50 dataset reported in Refs.~\cite{DarkSide-50:2022qzh,DarkSide:2022knj}. The published ionization spectrum is used directly, whereas the event timestamps are not publicly available. The modulation component of this case study is therefore illustrative: we assume that the data are consistent with no daily modulation and construct the corresponding background-only isoangle distribution.

\subsection{Dataset and Model Assumptions}
\label{sec:darkside_dataset}

DarkSide-50 operated with a fiducial liquid-argon target mass of \SI{19.4}{kg}, yielding a total exposure of \SI{12.7}{tonne}\text{-}days, or approximately \SI{0.42}{tonne}\text{-}months, over 653 days. We choose this dataset because its observed ionization spectrum, background models, and systematic-uncertainty prescriptions are available in the published literature over the range $3\leq N_e\leq170$~\cite{DarkSide-50:2022qzh,DarkSide:2022knj,DarkSide-50:2025umf,GlobalArgonDarkMatter:2022ppc,DarkSide-20k:2024yfq}. We restrict the present analysis to $3\leq N_e\leq10$, where the sub-GeV DM signal is concentrated; see \Cref{fig:modulated_dR_dNe_Ar}.

To model the event rates under the null hypothesis, we consider the following background components~\cite{DarkSide-50:2022qzh,DarkSide-50:2025umf}:
\begin{itemize}
\item $^{39}\mathrm{Ar}$ and $^{85}\mathrm{Kr}$ decays within the time-projection chamber;
\item PMT radioactivity, including contributions from $^{232}\mathrm{Th}$, $^{238}\mathrm{U}$, $^{40}\mathrm{K}$, and other isotopes;
\item spurious electrons (SEs), which may arise from ionization electrons trapped on electronegative impurities and subsequently released.
\end{itemize}
The radiogenic components account for most of the measured rate at $N_e\geq4$, whereas the SE background dominates below $N_e=4$~\cite{DarkSide-50:2022qzh}. We extract the nominal radiogenic spectra from Ref.~\cite{DarkSide-50:2022qzh} and model the SE contribution following Ref.~\cite{GlobalArgonDarkMatter:2022ppc}; further details are provided in Appendix~\ref{app:se_background_modelling}. Subdominant contributions, including cryostat radioactivity, are neglected.

For each signal hypothesis
$(m_\chi,\sigma_{\chi n}^{\mathrm{SI}},\bar{\sigma}_e)$,
the signal-plus-background model is obtained by adding the predicted DM ionization spectrum to these background components. We assign a $15\%$ bin-to-bin uncorrelated uncertainty to the total background, intended to approximately account for both normalization uncertainties in the radiogenic components and shape uncertainties in the SE model. For comparison, the reported correlated normalization uncertainties on the $^{39}\mathrm{Ar}$, $^{85}\mathrm{Kr}$, and PMT backgrounds are $14\%$, $4.7\%$, and $11.5\%$, respectively~\cite{DarkSide-50:2022qzh}, while DarkSide-20k projections assume a $20\%$ uncertainty on the SE normalization~\cite{DarkSide-20k:2024yfq}.

Although the ionization spectrum is publicly available, the event timestamps are not. We therefore assume an Asimov background-only isoangle distribution corresponding to event times distributed uniformly throughout the exposure. After mapping these timestamps through \Cref{eq:isoangle_definition}, the resulting $\Theta$ distribution follows the DarkSide-50 location-dependent exposure $w(\Theta)$. Thus, the spectral analysis uses the observed DarkSide-50 counts, while the shape analysis assumes that no daily modulation was observed.

\begin{figure*}[t]
\includegraphics[width=\textwidth]{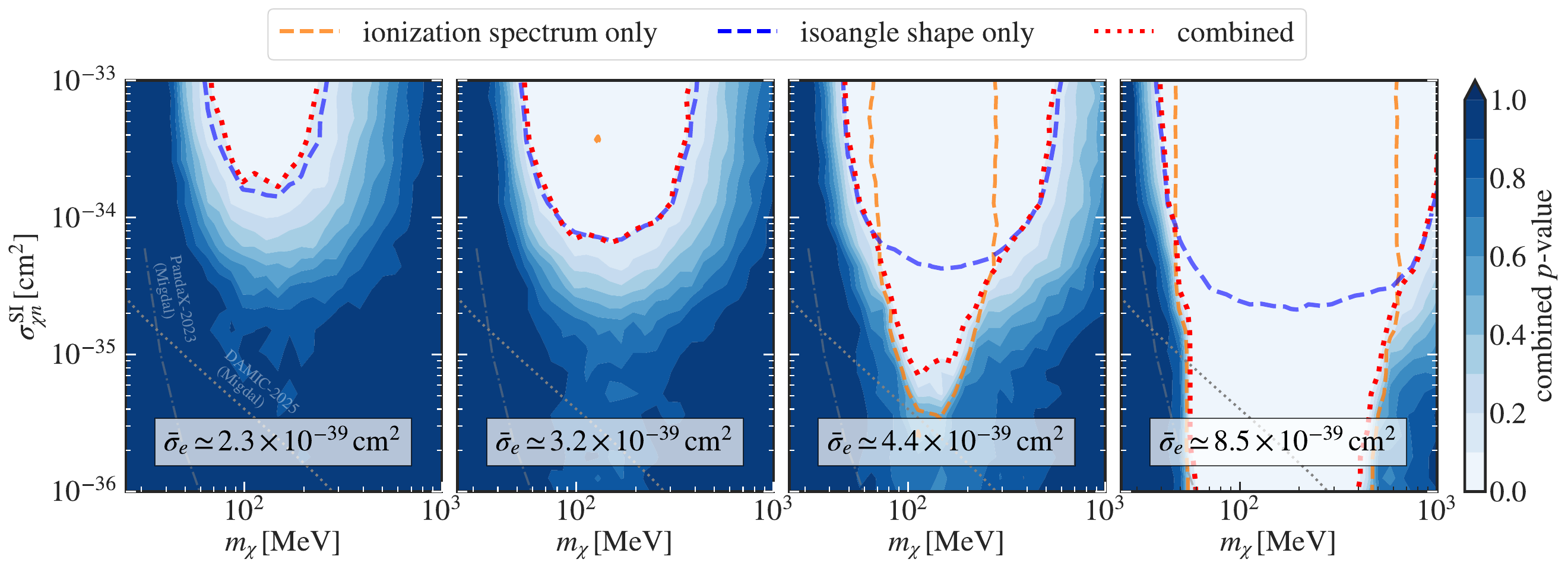}
\caption{Illustrative $90\%$ C.L. exclusions of selected
$(m_\chi,\sigma_{\chi n}^{\mathrm{SI}},\bar{\sigma}_e)$
values using the DarkSide-50 ionization spectrum and assuming that no daily modulation was observed. The dashed blue contour shows the isoangle shape constraint, defined by $p_{\mathrm{isoang}}=0.1$ in the $N_e=3$ bin. The dashed orange contour shows the spectral constraint, defined by $p_{\mathrm{spec}}=0.1$ using $3\leq N_e\leq10$. The dotted red contour shows the combined constraint, $p_{\mathrm{comb}}=0.1$, obtained using Fisher's method~\cite{Fisher:1925xxx,Mosteller:1948xxx}.}
\label{fig:darkside_KS_pvals_Ne3}
\end{figure*}

 \begin{figure}[t]
\includegraphics[width=\columnwidth]{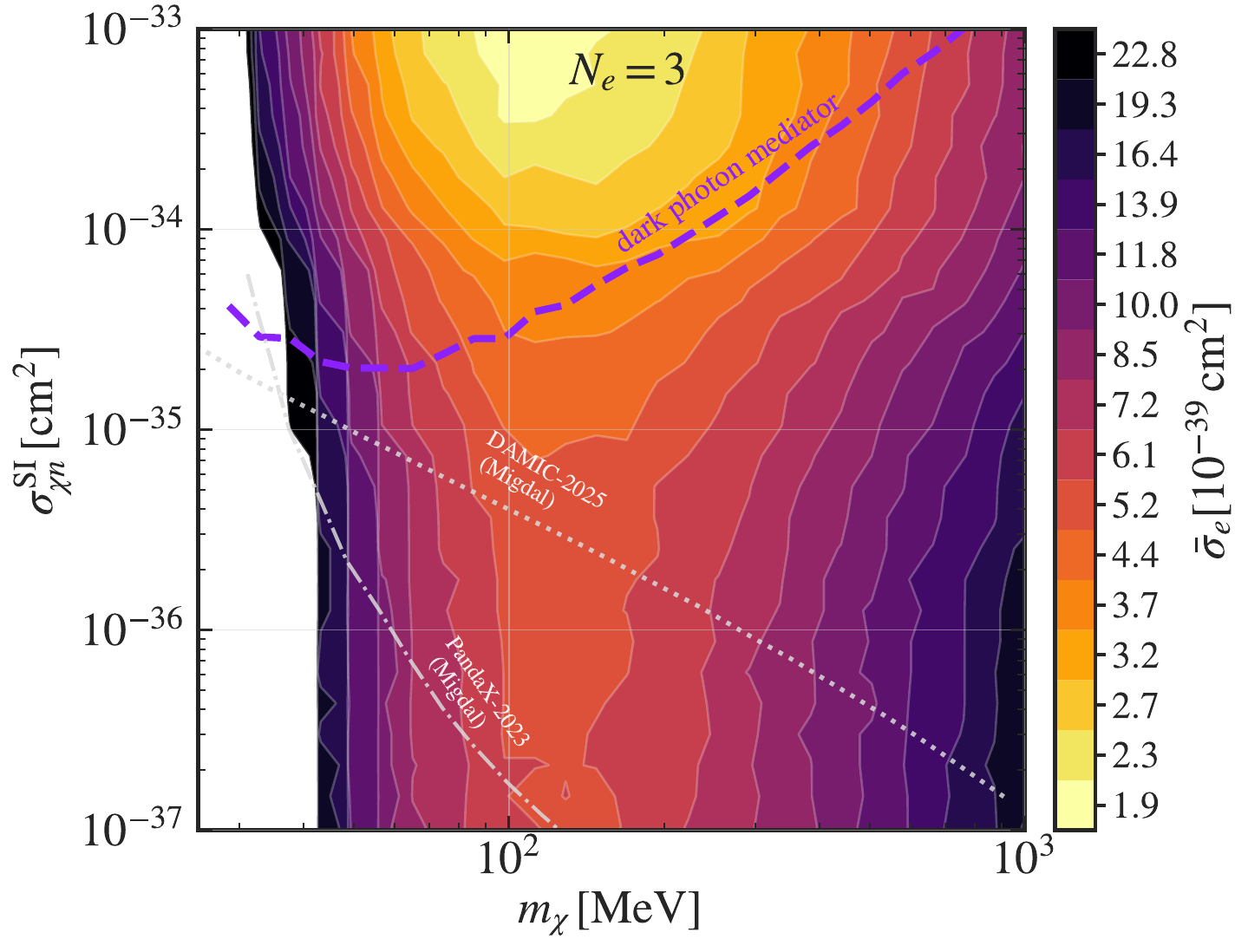}
\caption{Illustrative $90\%$ C.L. constraints on
$\sigma_{\chi n}^{\mathrm{SI}}$
as a function of DM mass for fixed values of $\bar{\sigma}_e$, obtained by combining the DarkSide-50 ionization-spectrum and isoangle-shape analyses. The filled contours are constructed in the same manner as the dotted red contours in \Cref{fig:darkside_KS_pvals_Ne3}. The purple dashed curve shows the corresponding constraint for the heavy-dark-photon mediator model.}
\label{fig:darkside_projected_constraints_3Ne}
\end{figure}

\subsection{Spectral and Isoangle Analyses}
\label{sec:darkside_analysis}

We test each signal hypothesis using two complementary observables: the ionization spectrum over $3\leq N_e\leq10$ and the isoangle distribution in the $N_e=3$ bin.

For the spectral analysis, we construct the signal-plus-background
$\mathrm{d}R/\mathrm{d}N_e$
template and fit it to the observed DarkSide-50 spectrum. The background pulls are varied to minimize the Poisson negative log-likelihood (NLL), including the assumed binwise systematic uncertainties. We perform an analogous fit under the background-only hypothesis and define
\begin{equation}
q_{\mathrm{spec}}
\equiv
\mathrm{NLL}_{\mathrm{sb}}^{\mathrm{min}}
-
\mathrm{NLL}_{\mathrm{b}}^{\mathrm{min}}.
\label{eq:darkside_spectral_statistic}
\end{equation}
Larger values of $q_{\mathrm{spec}}$ indicate that the tested signal-plus-background hypothesis provides a poorer description of the data relative to the background-only model.

To calibrate this statistic, we generate Poisson-fluctuated realizations of the tested signal-plus-background hypothesis at the best-fit background pulls and refit each realization using the same procedure. The spectral $p$-value,
$p_{\mathrm{spec}}\equiv p_{\mathrm{d}R/\mathrm{d}N_e}$,
is the fraction of realizations with a value of $q_{\mathrm{spec}}$ at least as large as that obtained from the observed spectrum.

For the modulation analysis, we apply the isoangle shape test introduced in \Cref{sec:stat_analysis} to the $N_e=3$ bin. The assumed data follow the unmodulated background-only $\Theta$ distribution, while each signal hypothesis predicts a distorted distribution due to Earth shielding. As before, we denote the corresponding CL$_\mathrm{s}$ $p$-value by $p_{\mathrm{isoang}}$. Because the assumed isoangle dataset is the median background-only expectation, we take $p_{\mathrm{b}}=0.5$ in the CL$_\mathrm{s}$ construction of \Cref{eq:cls_pb}.

The spectrum and isoangle shape carry complementary information. The spectral test uses the total event counts across $N_e$ bins, whereas the shape test uses the normalized distribution in $\Theta$ at fixed $N_e=3$. For a Poisson process, the total number of events and their distribution in $\Theta$ are statistically independent. We therefore combine $p_{\mathrm{spec}}$ and $p_{\mathrm{isoang}}$ using Fisher's method~\cite{Fisher:1925xxx,Mosteller:1948xxx}, where the test statistic $X_\mathrm{F}$ is defined as
\begin{equation}
X_{\mathrm{F}}
=
-2\left(
\ln p_{\mathrm{spec}}
+
\ln p_{\mathrm{isoang}}
\right)
\label{eq:darkside_fisher_statistic}
\end{equation}
 and converted to the combined $p$-value $p_{\mathrm{comb}}$ using a chi-squared distribution with four degrees of freedom.

\subsection{Combined Constraints}
\label{sec:darkside_constraints}

Figure~\ref{fig:darkside_KS_pvals_Ne3} compares the constraints obtained from the ionization spectrum, the $N_e=3$ isoangle shape, and their combination. The spectrum and shape tests probe different aspects of the signal: the former tests its energy dependence and normalization, while the latter tests the daily Earth-shielding pattern. Their combination consequently provides stronger constraints than either observable alone.

Figure~\ref{fig:darkside_projected_constraints_3Ne} shows the combined constraints over a wider range of $\bar{\sigma}_e$, together with the benchmark heavy-dark-photon model. Increasing $\bar{\sigma}_e$ raises the detector event rate and strengthens the sensitivity to Earth shielding until either $\sigma_{\chi n}^{\mathrm{SI}}$ is too small to produce an observable modulation signature or $\bar{\sigma}_e$ is too large to be compatible with the observed ionization spectrum.

Under the assumed absence of daily modulation, the resulting DarkSide-50 constraints lie largely within parameter space already excluded by Migdal-effect searches from PandaX and DAMIC-M~\cite{PandaX:2023xgl,DAMIC-M:2025luv}. This is expected from the smaller DarkSide-50 exposure, the restriction to $N_e\geq3$, and the inclusion of experimental systematic uncertainties, in contrast to the lower-threshold and larger-exposure projections shown in \Cref{fig:projections_Ne_1,fig:projections_Ne_2}.

Although not competitive with the strongest existing limits in this parameter space, which are Migdal-effect limits, this case study probes a physically distinct signal: DM--nucleus scattering during propagation through the Earth followed by DM--electron scattering in the detector. Migdal searches instead probe electronic excitation or ionization induced directly by a DM--nucleus collision in the detector, and have additional associated uncertainties. The DarkSide-50 example demonstrates how the spectral and daily-modulation framework developed here could be applied to timestamped data from larger current and future liquid-noble experiments.

\section{Summary and Outlook}

Daily modulation from Earth shielding provides a distinctive handle on sub-GeV DM interactions. In this work, we developed a framework in which DM couples to both electrons and nuclei: nuclear scattering in the Earth modifies the incident DM flux, while electron scattering in the detector produces the observable ionization signal. This separation makes the daily modulation pattern sensitive to the DM--nucleon interaction, while the overall ionization rate is controlled by the DM--electron interaction. The resulting signal therefore carries information that is not available in a rate-only search.

We quantified this effect for liquid-argon and liquid-xenon targets using Earth-propagated DM speed distributions from \texttt{DaMaSCUS}. We showed that the degree of attenuation depends strongly on the isoangle $\Theta$, which varies over a sidereal day as the detector rotates relative to the apparent DM wind. This produces a location-dependent modulation of the ionization spectrum. Detectors in the Northern and Southern Hemispheres can sample different ranges of $\Theta$, making laboratory location an important part of the signal interpretation. We further demonstrated that the DM--electron and DM--nucleon cross sections affect the observable signal in complementary ways: $\bar{\sigma}_e$ primarily sets the total event rate, while $\sigma_{\chi n}^{\mathrm{SI}}$ controls the strength and shape of the Earth-shielding modulation.

To exploit this information, we introduced a statistical analysis based on the shape of the observed $\Theta$ distribution, and showed how this can be combined with the ionization spectrum. This provides a way to test whether an excess has the sidereal structure expected from Earth shielding, rather than only whether it has the correct total rate. As a concrete example, we applied the method to DarkSide-50. Since public DarkSide-50 data do not include event timing information, we assumed no observed daily modulation and used the measured ionization spectrum together with the expected null $\Theta$ distribution. This illustrates how existing low-threshold data can be interpreted in the combined $(m_{\chi}, \sigma_{\chi n}^{\mathrm{SI}}, \bar{\sigma}_e)$ parameter space.

The broader implication is that Earth shielding can act as both a discovery channel and a validation tool. A sidereal modulation with the predicted phase, shape, target dependence, and site dependence would be difficult for many detector backgrounds to mimic, especially when compared across experiments. Conversely, the absence of such modulation can constrain scenarios in which nuclear scattering in the Earth is large enough to reshape the incident DM flux. Future low-threshold liquid-noble experiments, together with semiconductor and scintillator detectors, could therefore use timing information not only to improve sensitivity but also to diagnose the underlying interaction structure of a candidate signal.

Several extensions would further sharpen this program. A full treatment of annual variations in the Earth’s orbital velocity could be incorporated alongside the sidereal modulation considered here, allowing the combined daily and annual signal morphology to be used in searches. More detailed detector-specific studies, including time-dependent backgrounds, thresholds, efficiencies, and fewer-electron systematics, would make the method directly applicable to future data releases. Finally, applying the same framework across multiple detector materials and laboratory locations would provide a powerful cross-check of any candidate signal. Earth-shielding modulation thus opens a path toward using the Earth itself as part of the detector: not only attenuating the DM flux, but encoding information about the particle interactions responsible for it.

\section*{Acknowledgments}

We thank Jocelyn Monroe and Sergey Burdin for helpful discussions. JS and TK acknowledge support from the UK Research and Innovation Future Leader Fellowship~MR/Y018656/1.  RKL is supported by the U.S. Department of Energy under Contract DE-AC02-76SF00515.

\appendix

\counterwithin{figure}{section}
\setcounter{table}{0}
\renewcommand{\thetable}{A\arabic{table}}

\section{Model Space}
\label{app:models}
The DM model landscape is vast and a strong focus on any one candidate is not, in general, justified. Nevertheless, explicit models are highly instructive when assessing whether a given combination of signatures is plausible. In the present context, the relevant question is whether the interaction strengths governing DM--electron and DM--nucleon scattering are expected to be correlated, or whether they should be treated as independent effective parameters.

A simple example of correlated couplings is provided by a heavy dark-photon mediator. Let $\chi$ denote Dirac DM charged under a broken $U(1)_D$ with gauge coupling $g_D$, and let the corresponding gauge boson $A'$ kinetically mix with the photon with strength $\epsilon$. At momentum transfers well below the mediator mass, $q^2 \ll m_{A'}^2$, one obtains the effective interaction
\begin{equation}
\mathcal{L}_{\rm eff}
=
\frac{g_D \epsilon e}{m_{A'}^2}
\left(\bar{\chi}\gamma_\mu \chi\right) J^\mu_{\rm EM},
\end{equation}
where
\begin{equation}
J^\mu_{\rm EM} = \sum_f Q_f \, \bar f \gamma^\mu f.
\end{equation}
Defining
\begin{equation}
G_D \equiv \frac{g_D \epsilon e}{m_{A'}^2},
\end{equation}
the couplings to electrons and protons are fixed by the same coefficient,
\begin{equation}
\mathcal{L}_{\rm eff}
\supset
- G_D \left(\bar{\chi}\gamma_\mu \chi\right)\left(\bar e \gamma^\mu e\right)
+ G_D \left(\bar{\chi}\gamma_\mu \chi\right)\left(\bar p \gamma^\mu p\right),
\end{equation}
while the tree-level coupling to neutrons vanishes up to subleading effects. In the non-relativistic limit this implies spin-independent scattering cross sections
\begin{equation}
\sigma_{\chi e} \simeq \frac{\mu_{\chi e}^2}{\pi} G_D^2,
\qquad
\sigma_{\chi p} \simeq \frac{\mu_{\chi p}^2}{\pi} G_D^2,
\end{equation}
and therefore
\begin{equation}
\frac{\sigma_{\chi e}}{\sigma_{\chi p}}
\simeq
\frac{\mu_{\chi e}^2}{\mu_{\chi p}^2}.
\label{eq:dark_photon_xsec_relation}
\end{equation}
For a nucleus $N$ with charge $Z$, one correspondingly finds
\begin{equation}
\sigma_{\chi N} \simeq \frac{\mu_{\chi N}^2}{\pi} \, Z^2 G_D^2 \, F_N^2(q),
\end{equation}
such that the relative size of the electron and nuclear recoil rates is not arbitrary, but fixed up to reduced-mass effects, coherence factors, and form factors. In this sense, a heavy dark-photon mediator naturally correlates electron and nuclear scattering.

This should be contrasted with scenarios in which the couplings to leptons and baryons arise from distinct gauge interactions. A well-motivated example is furnished by theories where baryon and lepton number are separately gauged, $U(1)_B \otimes U(1)_L$ \cite{FileviezPerezWise2010}. In the Standard Model, this gauge structure is anomalous. In particular, one finds the non-vanishing mixed anomalies
\begin{equation}
\mathcal{A}\!\left[SU(2)^2 U(1)_B\right] = \frac{3}{2},
\qquad
\mathcal{A}\!\left[U(1)_Y^2 U(1)_B\right] = -\frac{3}{2},
\end{equation}
and analogously in the leptonic sector
\begin{equation}
\mathcal{A}\!\left[SU(2)^2 U(1)_L\right] = \frac{3}{2},
\qquad
\mathcal{A}\!\left[U(1)_Y^2 U(1)_L\right] = -\frac{3}{2}.
\end{equation}
The same reference shows that these anomalies can be cancelled by adding a single extra fermionic generation with suitable baryon and lepton number assignments, for example with new quarks carrying $B=1$ and new leptons carrying $L=3$ in the opposite-chirality realization \cite{FileviezPerezWise2010}. Closely related anomaly-free realizations of local baryon number with vector-like quarks were studied in Ref.~\cite{Duerr:2017zbi}, where the baryonic anomaly cancellation condition can be written as
\begin{equation}
B_1 - B_2 = - \frac{1}{n_f}.
\end{equation}

From the point of view of DM phenomenology, the crucial implication is that one may have two distinct neutral mediators, denoted here by $Z_B$ and $Z_L$, coupling separately to the baryon and lepton currents,
\begin{equation}
\mathcal{L}
\supset
g_B Z_{B\mu} J_B^\mu
+
g_L Z_{L\mu} J_L^\mu
+
g_\chi^B Z_{B\mu}\bar\chi\gamma^\mu\chi
+
g_\chi^L Z_{L\mu}\bar\chi\gamma^\mu\chi.
\end{equation}
In the heavy-mediator limit, integrating out both vectors yields
\begin{equation}
\mathcal{L}_{\rm eff}
\supset
\frac{g_\chi^B g_B}{m_{Z_B}^2}
(\bar\chi\gamma_\mu\chi) J_B^\mu
+
\frac{g_\chi^L g_L}{m_{Z_L}^2}
(\bar\chi\gamma_\mu\chi) J_L^\mu.
\end{equation}
Since electrons carry lepton number but no baryon number, whereas nucleons carry baryon number but no lepton number, one obtains at leading order
\begin{equation}
\sigma_{\chi e}
\simeq
\frac{\mu_{\chi e}^2}{\pi}
\left(
\frac{g_\chi^L g_L}{m_{Z_L}^2}
\right)^2,
\qquad
\sigma_{\chi p}
\simeq
\frac{\mu_{\chi p}^2}{\pi}
\left(
\frac{g_\chi^B g_B}{m_{Z_B}^2}
\right)^2,
\end{equation}
and, for coherent scattering on nuclei,
\begin{equation}
\sigma_{\chi N}
\simeq
\frac{\mu_{\chi N}^2}{\pi}
A^2
\left(
\frac{g_\chi^B g_B}{m_{Z_B}^2}
\right)^2
F_N^2(q).
\end{equation}
Consequently,
\begin{equation}
\frac{\sigma_{\chi e}}{\sigma_{\chi p}}
\simeq
\frac{\mu_{\chi e}^2}{\mu_{\chi p}^2}
\left(
\frac{g_\chi^L g_L}{g_\chi^B g_B}
\right)^2
\left(
\frac{m_{Z_B}}{m_{Z_L}}
\right)^4,
\end{equation}
which is not fixed by symmetry alone and can naturally span a wide range. Therefore, while single-mediator constructions such as the dark-photon portal predict correlated electron and nuclear recoil responses, multi-mediator theories based on gauged baryon and lepton number allow $\sigma_{\chi e}$ and $\sigma_{\chi N}$ to vary independently. This motivates treating both possibilities on equal footing in the interpretation of DM modulation signals.

\section{Spurious electron (SE) background in DarkSide-50}\label{app:se_background_modelling}

As per Ref.~\cite{GlobalArgonDarkMatter:2022ppc}, the spurious electron background component in DarkSide-50 can be modeled as $k$ electrons trapped and released following a primary ionization event. These secondary electrons may get piled up within a single time window with a pile-up probability $p_{\mathrm{SE}}$, resulting in a probability $P(k,\,p_{\mathrm{SE}})$ of them being reconstructed into a single S2 pulse:

\begin{equation}
    P(k,\,p_{\mathrm{SE}}) = \frac{1}{k!}\left(\frac{p_{\mathrm{SE}}^k}{k + 1} - \frac{p_{\mathrm{SE}}^{k+1}}{k + 2}\right).
\end{equation}
Given finite reconstruction resolution, the total of $k+1$ electrons (including the primary electron) are subject to Gaussian smearing $G\left(N_e\,|\,\mu,\,\sigma\right)$ with
\begin{equation}
    \mu = k+1;\,\sigma = F_{\mathrm{SE}}\sqrt{\frac{k+1}{g_2}},
\end{equation}
where $N_e$ is the number of reconstructed electrons, $g_2$ is the S2 gain factor ($23\,\mathrm{PE}\,/\,e^{-}$), and $F_{\mathrm{SE}}$ is the scale factor for the Gaussian width. The total SE rate in the reconstructed $N_e$ space then reads:
\begin{equation}
\label{eq:se_model}
    B_{\mathrm{SE}}(N_e) = R_{\mathrm{SE}} \sum_{k=0} P(k,\,p_{\rm SE}) G\left(N_e\,|\,\mu,\,\sigma\right),
\end{equation}
where $R_{\mathrm{SE}}$ is the SE spectrum normalization in events per kg-day. Combining the SE contribution with the $^{39}$Ar, $^{85}$Kr, and PMT backgrounds and their associated normalization uncertainties \cite{DarkSide-50:2022qzh}, we get the total background model:
\begin{align}
    B(N_e) &= B_{\mathrm{SE}}(N_e\,|\,R_{\mathrm{SE}},\,p_{\mathrm{SE}},\,F_{\mathrm{SE}}) \nonumber \\ &+ \sum_{\substack{i \in [\mathrm{PMT},\\\,^{39}\mathrm{Ar},^{85}\mathrm{Kr}]}}B_{i}(N_e\,|\,\sigma_{i}),  
\end{align}
where $\sigma_i$ are the Gaussian pulls of the radiogenic component normalizations. 

\begin{figure}[t]
\includegraphics[width=0.45\textwidth]{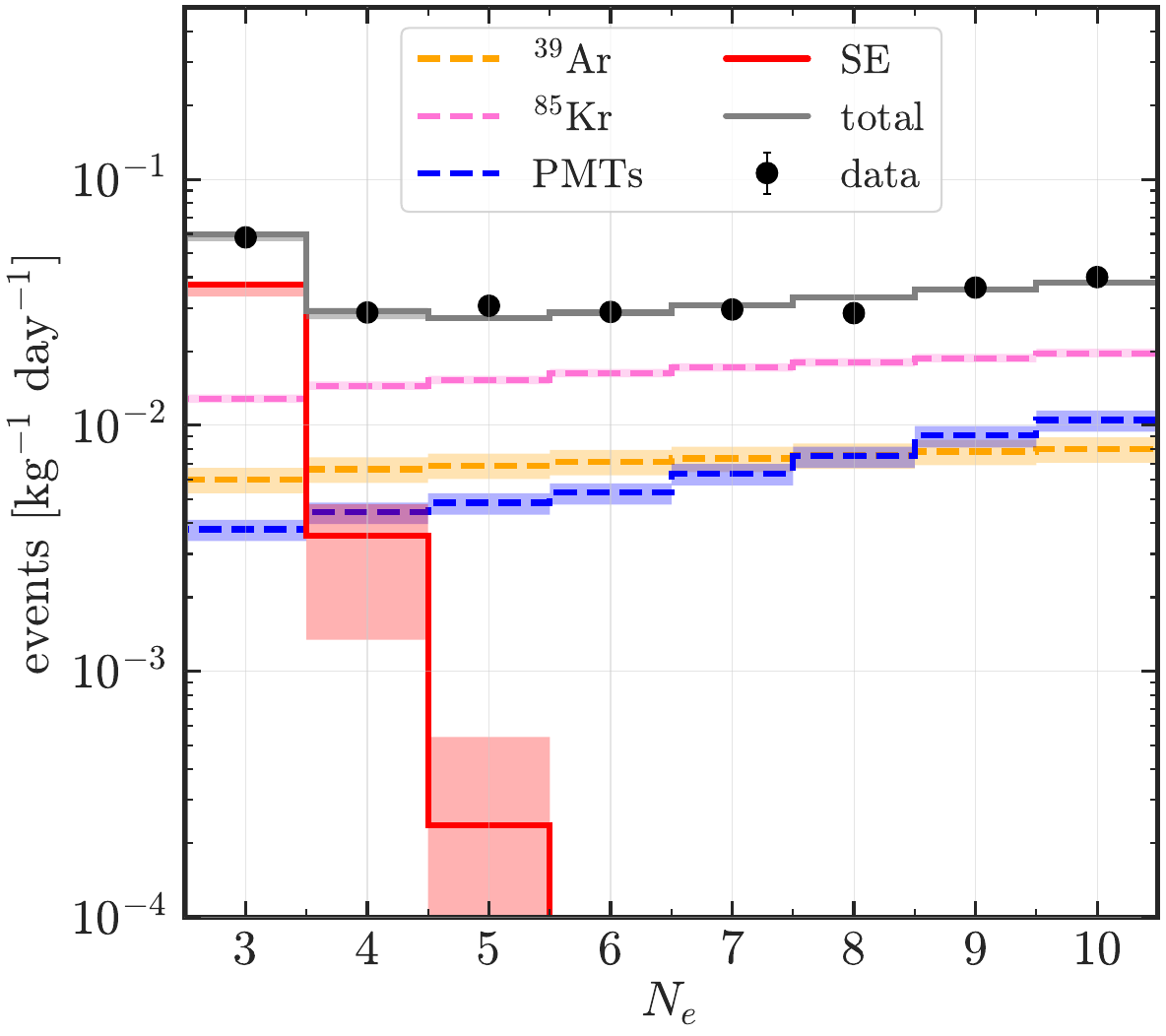}
\caption{Best-fitting backgrounds for 653 days of the DarkSide-50 data \cite{DarkSide-50:2022qzh}, re-binned here in $1\,N_e$-wide bins. The shaded bands represent the $68\%$ credible intervals derived from the \texttt{Ultranest} posterior samples. The spurious electron background model is given in \Cref{eq:se_model} (see also \cite{DarkSide-50:2025umf}).}
\label{fig:ds50_best_fit_backgrounds}
\end{figure}

Figure~\ref{fig:ds50_best_fit_backgrounds} shows the observed DarkSide-50 ionization spectrum along with the best-fitting total background and its individual components. The fit is performed to the DarkSide-50 data in the $3 \leq N_e \leq 10$ range using the Bayesian nested sampling approach via \texttt{Ultranest} \cite{Buchner:2021cql,Buchner:2024ult}. The maximum a posteriori (MAP) estimates of the fitted parameters are
\begin{align}
    &\hat{R}_{\mathrm{SE}} = 4.07;\,\hat{p}_{\mathrm{SE}} = 0.21;\,\hat{F}_{\mathrm{SE}} = 1.37; \label{eq:darkside_background_map_values}\\
    &\hat{\sigma}_{\mathrm{^{39}Ar}} = 0.17;\,\hat{\sigma}_{\mathrm{^{85}Kr}} = 0.10;\,\hat{\sigma}_{\mathrm{PMT}} = 0.23.
\end{align}
We use the SE MAP values from \Cref{eq:darkside_background_map_values} for the nominal spurious electron background model used in \Cref{sec:darkside_case}.

\section{Comment on Light Mediator Modulation}
\label{app:light_mediator_modulation}
The modulation signal for a light mediator is significantly different because the scattering, energy-loss, and diffusion processes exhibit strong momentum dependence. This leads to a much weaker effect of the Earth's shadow at comparable reference cross sections, as we discuss below.
For the light-mediator benchmark, the reduced modulation at a fixed quoted
DM--nucleon cross section is a physical consequence of both the cross-section
normalization convention and the momentum-transfer structure of the
interaction.  The implementation used in \texttt{DaMaSCUS} corresponds to an effectively
massless mediator coupled to electromagnetic charge, as realised, for example,
by a kinetically mixed dark photon.  The reference proton cross section
$\bar\sigma_{\chi p}$ is defined at the conventional momentum
\begin{equation}
    q_{\rm ref}=\alpha m_e,
\end{equation}
and should not be interpreted as the total zero-momentum cross section.  For a
neutral terrestrial atom of nuclear charge $Z$, the differential elastic cross
section is schematically
\begin{equation}
    \frac{\dd \sigma_A}{\dd q^2}
    =
    \frac{\sigma_{\chi p} Z^2}
         {4\mu_{\chi p}^2 v^2}
    \left(\frac{q_{\rm ref}}{q}\right)^4
    \left[
        \frac{a_A^2 q^2}{1+a_A^2q^2}
    \right]^2,
    \label{eq:screened_light_mediator_differential}
\end{equation}
where $a_A$ is the atomic screening radius.  The first momentum-dependent
factor is the light-mediator propagator, whereas the second is the atomic
charge-screening form factor.  At wavelengths larger than the atom, the
incident DM resolves the electrically neutral atom rather than the positively
charged nucleus alone.  Consequently, the electronic cloud cancels the
nuclear charge as $q\rightarrow0$, removing the otherwise divergent
Rutherford enhancement.

Using $q_{\max}=2\mu_{\chi A}v$, integration of
Eq.~\eqref{eq:screened_light_mediator_differential} gives
\begin{equation}
    \sigma_A(v)
    =
    \bar\sigma_{\chi p} Z^2
    \frac{\mu_{\chi A}^2}{\mu_{\chi p}^2}
    \frac{(a_A q_{\rm ref})^4}
         {1+4a_A^2\mu_{\chi A}^2v^2}.
    \label{eq:screened_light_mediator_total}
\end{equation}
Thus, even before considering the angular distribution of the collisions, the
actual atomic scattering probability can be substantially smaller than the
value suggested by the nominal reference cross section.  For the
Thomas--Fermi estimate $a_A\simeq0.885\,a_0 Z^{-1/3}$ and
$a_0^{-1}=\alpha m_e=q_{\rm ref}$, the low-velocity screening factor is
\begin{equation}
    (a_Aq_{\rm ref})^4
    \simeq
    0.61\,Z^{-4/3}.
    \label{eq:screening_suppression_scaling}
\end{equation}
The apparent coherent enhancement is therefore reduced from the unscreened
$Z^2$ scaling to approximately $Z^{2/3}$ in the saturated low-momentum limit.
A comparatively large value of $\sigma_{\chi p}$ may consequently be
required before the mean number of relevant collisions along an Earth-crossing
trajectory becomes appreciable.

More importantly, Earth shielding is controlled not only by the total number
of collisions, but by their ability to change the DM energy and direction.
Long-range scattering is strongly weighted towards small momentum transfer and
small scattering angle.  The quantities relevant for stopping and directional
redistribution are therefore transport-weighted cross sections, for example
\begin{equation}
    \sigma_T(v)
    =
    \int_0^{q_{\max}^2}\dd q^2\,
    \frac{q^2}{2\mu_{\chi A}^2v^2}
    \frac{\dd\sigma_A}{\dd q^2},
    \label{eq:transport_cross_section}
\end{equation}
with an analogous energy-loss weighting for the stopping power.  The factor
$q^2/(2\mu_{\chi A}^2v^2)=1-\cos\theta$ suppresses the contribution of the
numerous forward scatters.  A light mediator can therefore generate a sizable
raw collision rate while producing only modest cumulative deflection or
energy loss.  By contrast, a contact interaction distributes events much more
evenly over the kinematically allowed recoil range and transfers a larger
fraction of the incident momentum per collision.

The daily modulation is generated by the difference between short and long
terrestrial trajectories as the isoangle $\Theta$ changes during a sidereal
day.  A useful schematic measure is the transport optical depth,
\begin{equation}
    \tau_T(\Theta,v)
    =
    \sum_A\int_{\mathrm{path}(\Theta)}\dd \ell\,
    n_A(\ell)\,\sigma_{T,A}(v),
    \label{eq:transport_optical_depth}
\end{equation}
which controls the accumulated angular diffusion and is more directly related
to modulation than the optical depth formed from the total cross section.
When $\tau_T\ll1$ for both the short and long trajectories, the incident
velocity distributions remain close to the Standard Halo Model and their
difference is small.  Atomic screening and forward-peaked scattering both
reduce $\tau_T$ at fixed $\bar\sigma_{\chi p}$.  The light-mediator scenario
therefore reaches the regime of pronounced path-length-dependent attenuation
only at substantially larger reference cross sections than the contact case.
Equivalently, at the same quoted reference cross section, the contrast between
the small-$\Theta$ and large-$\Theta$ fluxes is weaker, leading to a smaller
daily modulation amplitude.

This comparison should be understood at fixed \emph{reference} cross section.
Because $\bar\sigma_{\chi p}$ is defined at $q_{\rm ref}$, its numerical value
is not a model-independent measure of the integrated scattering probability.
The displacement of the modulation feature towards larger
$\bar\sigma_{\chi p}$ is therefore partly a consequence of this convention,
but the reduced transport efficiency is physical.  Finally, the screening
factor in Eq.~\eqref{eq:screened_light_mediator_differential} is specific to a
mediator coupled to electromagnetic charge.  It would not apply unchanged to
an ultralight mediator coupled to baryon number, for which ordinary atoms are
not neutral under the new force.  The latter case would generally yield more
efficient Earth scattering at fixed nucleon reference cross section, but is
also subject to strong independent constraints on long-range baryonic forces.

%%%%%%%%%%%%%%%%%%%%%%%%%%%%%%%%%%%%%%%%%%%%%%%%%%%%%%%%%
%%%%%%%%%%%%%%%%%%%%%%%%%%%%%%%%%%%%%%%%%%%%%%%%%%%%%%%%%

\clearpage
\twocolumngrid
\bibliography{bibliography}

@article{PandaX:2023xgl,
    author = "Huang, Di and others",
    collaboration = "PandaX",
    title = "{Search for Dark-Matter{\textendash}Nucleon Interactions with a Dark Mediator in PandaX-4T}",
    eprint = "2308.01540",
    archivePrefix = "arXiv",
    primaryClass = "hep-ex",
    doi = "10.1103/PhysRevLett.131.191002",
    journal = "Phys. Rev. Lett.",
    volume = "131",
    number = "19",
    pages = "191002",
    year = "2023"
}

@article{GlobalArgonDarkMatter:2022ppc,
    author = "Agnes, P. and others",
    collaboration = "Global Argon Dark Matter",
    title = "{Sensitivity projections for a dual-phase argon TPC optimized for light dark matter searches through the ionization channel}",
    eprint = "2209.01177",
    archivePrefix = "arXiv",
    primaryClass = "physics.ins-det",
    doi = "10.1103/PhysRevD.107.112006",
    journal = "Phys. Rev. D",
    volume = "107",
    number = "11",
    pages = "112006",
    year = "2023"
}

@article{DarkSide-50:2022qzh,
    author = "Agnes, P. and others",
    collaboration = "DarkSide-50",
    title = "{Search for low-mass dark matter WIMPs with 12~ton-day exposure of DarkSide-50}",
    eprint = "2207.11966",
    archivePrefix = "arXiv",
    primaryClass = "hep-ex",
    reportNumber = "FERMILAB-PUB-22-589-ND-PPD-SCD",
    doi = "10.1103/PhysRevD.107.063001",
    journal = "Phys. Rev. D",
    volume = "107",
    number = "6",
    pages = "063001",
    year = "2023"
}

@article{DarkSide:2022knj,
    author = "Agnes, P. and others",
    collaboration = "DarkSide",
    title = "{Search for Dark Matter Particle Interactions with Electron Final States with DarkSide-50}",
    eprint = "2207.11968",
    archivePrefix = "arXiv",
    primaryClass = "hep-ex",
    reportNumber = "FERMILAB-PUB-22-566-ND-PPD-SCD",
    doi = "10.1103/PhysRevLett.130.101002",
    journal = "Phys. Rev. Lett.",
    volume = "130",
    number = "10",
    pages = "101002",
    year = "2023"
}

@article{DarkSide-50:2025umf,
    author = "Agnes, P. and others",
    collaboration = "DarkSide-50",
    title = "{Characterization of spurious-electron signals in the double-phase argon TPC of the DarkSide-50 experiment}",
    eprint = "2507.23003",
    archivePrefix = "arXiv",
    primaryClass = "hep-ex",
    reportNumber = "FERMILAB-PUB-25-0613-CSAID",
    month = "7",
    year = "2025",
    journal = "" 
}

@article{DarkSide-20k:2024yfq,
    author = "Acerbi, F. and others",
    collaboration = "DarkSide-20k",
    title = "{DarkSide-20k sensitivity to light dark matter particles}",
    eprint = "2407.05813",
    archivePrefix = "arXiv",
    primaryClass = "hep-ex",
    reportNumber = "FERMILAB-PUB-24-0483-V",
    doi = "10.1038/s42005-024-01896-z",
    journal = "Commun. Phys.",
    volume = "7",
    number = "1",
    pages = "422",
    year = "2024"
}

@misc{Buchner:2024ult,
  url = {https://johannesbuchner.github.io/UltraNest/},
  author = {Buchner, Johannes},
  title = "{UltraNest Documentation}",
  year = {2024},
  note={\href{https://johannesbuchner.github.io/UltraNest/}{https://johannesbuchner.github.io/UltraNest/}}
}

@article{Buchner:2021cql,
    author = "Buchner, Johannes",
    title = "{UltraNest -- A Robust, General Purpose Bayesian Inference Engine}",
    eprint = "2101.09604",
    archivePrefix = "arXiv",
    primaryClass = "stat.CO",
    reportNumber = "10.21105/joss.03001",
    month = "1",
    year = "2021",
    journal = ""
}

@article{FileviezPerezWise2010,
  author        = {Fileviez P{\'e}rez, Pavel and Wise, Mark B.},
  title         = {Baryon and Lepton Number as Local Gauge Symmetries},
  eprint        = {1002.1754},
  archivePrefix = {arXiv},
  primaryClass  = {hep-ph},
  year          = {2010},
  note          = {arXiv:1002.1754}
}

@article{Duerr:2017zbi,
  author        = {Duerr, Michael and Fileviez P{\'e}rez, Pavel and Smirnov, Juri},
  title         = {Baryonic Higgs at the LHC},
  eprint        = {1704.03811},
  archivePrefix = {arXiv},
  primaryClass  = {hep-ph},
  reportNumber  = {DESY-17-050},
  year          = {2017},
  note          = {arXiv:1704.03811}
}

@article{Baxter:2021pqo,
    author = "Baxter, D. and others",
    title = "{Recommended conventions for reporting results from direct dark matter searches}",
    eprint = "2105.00599",
    archivePrefix = "arXiv",
    primaryClass = "hep-ex",
    doi = "10.1140/epjc/s10052-021-09655-y",
    journal = "Eur. Phys. J. C",
    volume = "81",
    number = "10",
    pages = "907",
    year = "2021"
}

@article{SENSEI:2025qvp,
    author = "Bloch, Itay M. and others",
    collaboration = "SENSEI",
    title = "{SENSEI: A Search for Diurnal Modulation in sub-GeV Dark Matter Scattering}",
    eprint = "2510.20889",
    archivePrefix = "arXiv",
    primaryClass = "hep-ex",
    reportNumber = "FERMILAB-PUB-25-0785-CSAID-PPD",
    month = "10",
    year = "2025"
}

@article{DAMIC-M:2025ltz,
    author = "Aggarwal, K. and others",
    collaboration = "DAMIC-M",
    title = "{Daily Modulation Constraints on Light Dark Matter with DAMIC-M}",
    eprint = "2511.13962",
    archivePrefix = "arXiv",
    primaryClass = "hep-ex",
    month = "11",
    year = "2025"
}

@article{QUEST-DMC:2025qsa,
    author = "Darvishi, N. and others",
    collaboration = "QUEST-DMC",
    title = "{Dark matter attenuation effects: sensitivity ceilings for spin-dependent and spin-independent interactions}",
    eprint = "2502.10251",
    archivePrefix = "arXiv",
    primaryClass = "hep-ph",
    doi = "10.1088/1475-7516/2025/04/017",
    journal = "JCAP",
    volume = "04",
    pages = "017",
    year = "2025"
}

@article{Bertou:2025adb,
    author = "Bertou, Xavier and Desai, Ansh and Emken, Timon and Essig, Rouven and Volansky, Tomer and Yu, Tien-Tien",
    title = "{Earth-scattering induced modulation in low-threshold dark matter experiments}",
    eprint = "2507.00344",
    archivePrefix = "arXiv",
    primaryClass = "hep-ph",
    doi = "10.1007/JHEP11(2025)042",
    journal = "JHEP",
    volume = "11",
    pages = "042",
    year = "2025"
}

@article{Cappiello:2023hza,
    author = "Cappiello, Christopher V.",
    title = "{Analytic Approach to Light Dark Matter Propagation}",
    eprint = "2301.07728",
    archivePrefix = "arXiv",
    primaryClass = "hep-ph",
    doi = "10.1103/PhysRevLett.130.221001",
    journal = "Phys. Rev. Lett.",
    volume = "130",
    number = "22",
    pages = "221001",
    year = "2023"
}

@article{Emken:2018run,
    author = "Emken, Timon and Kouvaris, Chris",
    title = "{How blind are underground and surface detectors to strongly interacting Dark Matter?}",
    eprint = "1802.04764",
    archivePrefix = "arXiv",
    primaryClass = "hep-ph",
    reportNumber = "CP3-ORIGINS-2018-007, CP3-Origins-2018-007 DNRF90",
    doi = "10.1103/PhysRevD.97.115047",
    journal = "Phys. Rev. D",
    volume = "97",
    number = "11",
    pages = "115047",
    year = "2018"
}

@article{Bandyopadhyay:2010zj,
    author = "Bandyopadhyay, Abhijit and Majumdar, Debasish",
    title = "{On Diurnal and Annual Variations of Directional Detection Rates of Dark Matter}",
    eprint = "1006.3231",
    archivePrefix = "arXiv",
    primaryClass = "hep-ph",
    reportNumber = "SINP-APC-2010-01, RKMVU-PHY-2010-03",
    doi = "10.1088/0004-637X/746/1/107",
    journal = "Astrophys. J.",
    volume = "746",
    pages = "107",
    year = "2012"
}

@article{Essig:2011nj,
    author = "Essig, Rouven and Mardon, Jeremy and Volansky, Tomer",
    title = "{Direct Detection of Sub-GeV Dark Matter}",
    eprint = "1108.5383",
    archivePrefix = "arXiv",
    primaryClass = "hep-ph",
    reportNumber = "SLAC-PUB-14538",
    doi = "10.1103/PhysRevD.85.076007",
    journal = "Phys. Rev. D",
    volume = "85",
    pages = "076007",
    year = "2012"
}

@article{Essig:2017kqs,
    author = "Essig, Rouven and Volansky, Tomer and Yu, Tien-Tien",
    title = "{New Constraints and Prospects for sub-GeV Dark Matter Scattering off Electrons in Xenon}",
    eprint = "1703.00910",
    archivePrefix = "arXiv",
    primaryClass = "hep-ph",
    reportNumber = "CERN-TH-2017-042, YITP-SB-17-09",
    doi = "10.1103/PhysRevD.96.043017",
    journal = "Phys. Rev. D",
    volume = "96",
    number = "4",
    pages = "043017",
    year = "2017"
}

@article{Bunge:1993jsz,
    author = "Bunge, C. F. and Barrientos, J. A. and Bunge, A. V.",
    title = "{Roothaan-Hartree-Fock Ground-State Atomic Wave Functions: Slater-Type Orbital Expansions and Expectation Values for Z = 2-54}",
    doi = "10.1006/adnd.1993.1003",
    journal = "Atom. Data Nucl. Data Tabl.",
    volume = "53",
    pages = "113--162",
    year = "1993"
}

@article{Catena:2025ung,
    author = "Catena, Riccardo and Marin, Luca and Matas, Marek and Spaldin, Nicola A. and Urdshals, Einar",
    title = "{Electronic structure of liquid xenon in the context of light dark matter direct detection}",
    eprint = "2502.02965",
    archivePrefix = "arXiv",
    primaryClass = "hep-ph",
    doi = "10.21468/SciPostPhys.19.3.064",
    journal = "SciPost Phys.",
    volume = "19",
    number = "3",
    pages = "064",
    year = "2025"
}

@software{DarkArt:2022xxx,
  author       = {Timon Emken},
  title        = {DarkART: Version 0.1.0},
  month        = feb,
  year         = 2022,
  publisher    = {Zenodo},
  version      = {v0.1.0},
  doi          = {10.5281/zenodo.6046225},
  url          = {https://doi.org/10.5281/zenodo.6046225},
}

@software{DaMaSCUS:2020xxx,
  author       = {Timon Emken},
  title        = {temken/DaMaSCUS: Version 1.1},
  month        = mar,
  year         = 2020,
  publisher    = {Zenodo},
  version      = {v1.1},
  doi          = {10.5281/zenodo.3726878},
  url          = {https://doi.org/10.5281/zenodo.3726878},
}

@article{Blanco:2022cel,
    author = "Blanco, Carlos and Essig, Rouven and Fernandez-Serra, Marivi and Ramani, Harikrishnan and Slone, Oren",
    title = "{Dark matter direct detection with quantum dots}",
    eprint = "2208.05967",
    archivePrefix = "arXiv",
    primaryClass = "hep-ph",
    doi = "10.1103/PhysRevD.107.095035",
    journal = "Phys. Rev. D",
    volume = "107",
    number = "9",
    pages = "095035",
    year = "2023"
}

@article{Essig:2015cda,
    author = "Essig, Rouven and Fernandez-Serra, Marivi and Mardon, Jeremy and Soto, Adrian and Volansky, Tomer and Yu, Tien-Tien",
    title = "{Direct Detection of sub-GeV Dark Matter with Semiconductor Targets}",
    eprint = "1509.01598",
    archivePrefix = "arXiv",
    primaryClass = "hep-ph",
    doi = "10.1007/JHEP05(2016)046",
    journal = "JHEP",
    volume = "05",
    pages = "046",
    year = "2016"
}

@article{Blanco:2019lrf,
    author = "Blanco, Carlos and Collar, J. I. and Kahn, Yonatan and Lillard, Benjamin",
    title = "{Dark Matter-Electron Scattering from Aromatic Organic Targets}",
    eprint = "1912.02822",
    archivePrefix = "arXiv",
    primaryClass = "hep-ph",
    doi = "10.1103/PhysRevD.101.056001",
    journal = "Phys. Rev. D",
    volume = "101",
    number = "5",
    pages = "056001",
    year = "2020"
}

@inproceedings{Alfonso-Pita:2022akn,
    author = "Alfonso-Pita, E. and others",
    title = "{Snowmass 2021 Scintillating Bubble Chambers: Liquid-noble Bubble Chambers for Dark Matter and CE$\nu$NS Detection}",
    booktitle = "{Snowmass 2021}",
    eprint = "2207.12400",
    archivePrefix = "arXiv",
    primaryClass = "physics.ins-det",
    reportNumber = "FERMILAB-CONF-22-535-LDRD-PPD",
    month = "7",
    year = "2022"
}

@article{Krnjaic:2022ozp,
    author = "Krnjaic, G. and others",
    title = "{A Snowmass Whitepaper: Dark Matter Production at Intensity-Frontier Experiments}",
    eprint = "2207.00597",
    archivePrefix = "arXiv",
    primaryClass = "hep-ph",
    reportNumber = "FERMILAB-PUB-22-497-T",
    month = "7",
    year = "2022",
    journal=""
}

@inproceedings{SuperCDMS:2022kse,
    author = "Albakry, M. F. and others",
    collaboration = "SuperCDMS",
    title = "{A Strategy for Low-Mass Dark Matter Searches with Cryogenic Detectors in the SuperCDMS SNOLAB Facility}",
    booktitle = "{Snowmass 2021}",
    eprint = "2203.08463",
    archivePrefix = "arXiv",
    primaryClass = "physics.ins-det",
    reportNumber = "FERMILAB-CONF-22-171-PPD-SQMS",
    month = "3",
    year = "2022"
}

@inproceedings{Akesson:2022vza,
    author = "{\r{A}}kesson, Torsten and others",
    title = "{Current Status and Future Prospects for the Light Dark Matter eXperiment}",
    booktitle = "{Snowmass 2021}",
    eprint = "2203.08192",
    archivePrefix = "arXiv",
    primaryClass = "hep-ex",
    reportNumber = "FERMILAB-CONF-22-313-PPD-SCD-T",
    month = "3",
    year = "2022"
}

@article{Mitridate:2022tnv,
    author = "Mitridate, Andrea and Trickle, Tanner and Zhang, Zhengkang and Zurek, Kathryn M.",
    title = "{Snowmass white paper: Light dark matter direct detection at the interface with condensed matter physics}",
    eprint = "2203.07492",
    archivePrefix = "arXiv",
    primaryClass = "hep-ph",
    reportNumber = "FERMILAB-PUB-23-551-T",
    doi = "10.1016/j.dark.2023.101221",
    journal = "Phys. Dark Univ.",
    volume = "40",
    pages = "101221",
    year = "2023"
}

@article{Wang:2022cyk,
    author = "Wang, G. and Chang, C. L. and Lisovenko, M. and Novosad, V. and Yefremenko, V. G. and Zhang, J.",
    title = "{Light Dark Matter Detection with Hydrogen-Rich Targets and Low-$T_c$ TES Detectors}",
    eprint = "2201.04219",
    archivePrefix = "arXiv",
    primaryClass = "physics.ins-det",
    doi = "10.1007/s10909-022-02784-y",
    journal = "J. Low Temp. Phys.",
    volume = "209",
    number = "3-4",
    pages = "379--388",
    year = "2022"
}

@inproceedings{Battaglieri:2017aum,
    author = "Battaglieri, Marco and others",
    title = "{US Cosmic Visions: New Ideas in Dark Matter 2017: Community Report}",
    booktitle = "{U.S. Cosmic Visions: New Ideas in Dark Matter}",
    eprint = "1707.04591",
    archivePrefix = "arXiv",
    primaryClass = "hep-ph",
    reportNumber = "FERMILAB-CONF-17-282-AE-PPD-T",
    month = "7",
    year = "2017"
}

@article{DarkSide:2022dhx,
    author = "Agnes, P. and others",
    collaboration = "DarkSide",
    title = "{Search for Dark-Matter\textendash{}Nucleon Interactions via Migdal Effect with DarkSide-50}",
    eprint = "2207.11967",
    archivePrefix = "arXiv",
    primaryClass = "hep-ex",
    doi = "10.1103/PhysRevLett.130.101001",
    journal = "Phys. Rev. Lett.",
    volume = "130",
    number = "10",
    pages = "101001",
    year = "2023"
}

@article{LZ:2024zvo,
    author = "Aalbers, J. and others",
    collaboration = "LZ",
    title = "{Dark Matter Search Results from 4.2{\,}{\,}Tonne-Years of Exposure of the LUX-ZEPLIN (LZ) Experiment}",
    eprint = "2410.17036",
    archivePrefix = "arXiv",
    primaryClass = "hep-ex",
    reportNumber = "FERMILAB-PUB-24-0796-V",
    doi = "10.1103/4dyc-z8zf",
    journal = "Phys. Rev. Lett.",
    volume = "135",
    number = "1",
    pages = "011802",
    year = "2025"
}

@article{Blanco:2021hlm,
    author = "Blanco, Carlos and Kahn, Yonatan and Lillard, Benjamin and McDermott, Samuel D.",
    title = "{Dark Matter Daily Modulation With Anisotropic Organic Crystals}",
    eprint = "2103.08601",
    archivePrefix = "arXiv",
    primaryClass = "hep-ph",
    reportNumber = "FERMILAB-PUB-21-066-T",
    doi = "10.1103/PhysRevD.104.036011",
    journal = "Phys. Rev. D",
    volume = "104",
    pages = "036011",
    year = "2021"
}

@article{Migdal:1941xxx,
  author       = {Migdal, A. B.},
  title        = {Ionization of Atoms Accompanying Alpha and Beta Decay},
  journal      = {J. Phys. (USSR)},
  volume       = {4},
  pages        = {449--453},
  year         = {1941}
}

@article{Graham:2012su,
    author = "Graham, Peter W. and Kaplan, David E. and Rajendran, Surjeet and Walters, Matthew T.",
    title = "{Semiconductor Probes of Light Dark Matter}",
    eprint = "1203.2531",
    archivePrefix = "arXiv",
    primaryClass = "hep-ph",
    doi = "10.1016/j.dark.2012.09.001",
    journal = "Phys. Dark Univ.",
    volume = "1",
    pages = "32--49",
    year = "2012"
}

@article{Essig:2019xkx,
    author = "Essig, Rouven and Pradler, Josef and Sholapurkar, Mukul and Yu, Tien-Tien",
    title = "{Relation between the Migdal Effect and Dark Matter-Electron Scattering in Isolated Atoms and Semiconductors}",
    eprint = "1908.10881",
    archivePrefix = "arXiv",
    primaryClass = "hep-ph",
    reportNumber = "YITP-19-23",
    doi = "10.1103/PhysRevLett.124.021801",
    journal = "Phys. Rev. Lett.",
    volume = "124",
    number = "2",
    pages = "021801",
    year = "2020"
}

@article{Kurinsky:2019pgb,
    author = "Kurinsky, Noah Alexander and Yu, To Chin and Hochberg, Yonit and Cabrera, Blas",
    title = "{Diamond Detectors for Direct Detection of Sub-GeV Dark Matter}",
    eprint = "1901.07569",
    archivePrefix = "arXiv",
    primaryClass = "hep-ex",
    reportNumber = "FERMILAB-PUB-19-020-AE-E",
    doi = "10.1103/PhysRevD.99.123005",
    journal = "Phys. Rev. D",
    volume = "99",
    number = "12",
    pages = "123005",
    year = "2019"
}

@article{Griffin:2020lgd,
    author = "Griffin, Sin{\'e}ad M. and Hochberg, Yonit and Inzani, Katherine and Kurinsky, Noah and Lin, Tongyan and Chin, To",
    title = "{Silicon carbide detectors for sub-GeV dark matter}",
    eprint = "2008.08560",
    archivePrefix = "arXiv",
    primaryClass = "hep-ph",
    reportNumber = "FERMILAB-PUB-20-469-AE",
    doi = "10.1103/PhysRevD.103.075002",
    journal = "Phys. Rev. D",
    volume = "103",
    number = "7",
    pages = "075002",
    year = "2021"
}

@article{Hochberg:2015pha,
    author = "Hochberg, Yonit and Zhao, Yue and Zurek, Kathryn M.",
    title = "{Superconducting Detectors for Superlight Dark Matter}",
    eprint = "1504.07237",
    archivePrefix = "arXiv",
    primaryClass = "hep-ph",
    doi = "10.1103/PhysRevLett.116.011301",
    journal = "Phys. Rev. Lett.",
    volume = "116",
    number = "1",
    pages = "011301",
    year = "2016"
}

@article{Hochberg:2019cyy,
    author = "Hochberg, Yonit and Charaev, Ilya and Nam, Sae-Woo and Verma, Varun and Colangelo, Marco and Berggren, Karl K.",
    title = "{Detecting Sub-GeV Dark Matter with Superconducting Nanowires}",
    eprint = "1903.05101",
    archivePrefix = "arXiv",
    primaryClass = "hep-ph",
    doi = "10.1103/PhysRevLett.123.151802",
    journal = "Phys. Rev. Lett.",
    volume = "123",
    number = "15",
    pages = "151802",
    year = "2019"
}

@article{Hochberg:2021ymx,
    author = "Hochberg, Yonit and Kramer, Eric David and Kurinsky, Noah and Lehmann, Benjamin V.",
    title = "{Directional detection of light dark matter in superconductors}",
    eprint = "2109.04473",
    archivePrefix = "arXiv",
    primaryClass = "hep-ph",
    reportNumber = "FERMILAB-PUB-21-540-V",
    doi = "10.1103/PhysRevD.107.076015",
    journal = "Phys. Rev. D",
    volume = "107",
    number = "7",
    pages = "076015",
    year = "2023"
}

@article{Hochberg:2021yud,
    author = "Hochberg, Yonit and Lehmann, Benjamin V. and Charaev, Ilya and Chiles, Jeff and Colangelo, Marco and Nam, Sae Woo and Berggren, Karl K.",
    title = "{New constraints on dark matter from superconducting nanowires}",
    eprint = "2110.01586",
    archivePrefix = "arXiv",
    primaryClass = "hep-ph",
    doi = "10.1103/PhysRevD.106.112005",
    journal = "Phys. Rev. D",
    volume = "106",
    number = "11",
    pages = "112005",
    year = "2022"
}

@article{QROCODILE:2024zmg,
    author = "Baudis, Laura and others",
    collaboration = "QROCODILE",
    title = "{A New Bite Into Dark Matter with the SNSPD-Based QROCODILE Experiment}",
    eprint = "2412.16279",
    archivePrefix = "arXiv",
    primaryClass = "hep-ph",
    reportNumber = "MIT-CTP/5744",
    month = "12",
    year = "2024"
}

@article{Schutz:2016tid,
    author = "Schutz, Katelin and Zurek, Kathryn M.",
    title = "{Detectability of Light Dark Matter with Superfluid Helium}",
    eprint = "1604.08206",
    archivePrefix = "arXiv",
    primaryClass = "hep-ph",
    doi = "10.1103/PhysRevLett.117.121302",
    journal = "Phys. Rev. Lett.",
    volume = "117",
    number = "12",
    pages = "121302",
    year = "2016"
}

@article{Hochberg:2017wce,
    author = "Hochberg, Yonit and Kahn, Yonatan and Lisanti, Mariangela and Zurek, Kathryn M. and Grushin, Adolfo G. and Ilan, Roni and Griffin, Sin{\'e}ad M. and Liu, Zhen-Fei and Weber, Sophie F. and Neaton, Jeffrey B.",
    title = "{Detection of sub-MeV Dark Matter with Three-Dimensional Dirac Materials}",
    eprint = "1708.08929",
    archivePrefix = "arXiv",
    primaryClass = "hep-ph",
    reportNumber = "PUPT-2535",
    doi = "10.1103/PhysRevD.97.015004",
    journal = "Phys. Rev. D",
    volume = "97",
    number = "1",
    pages = "015004",
    year = "2018"
}

@article{Geilhufe:2019ndy,
    author = "Geilhufe, R. Matthias and Kahlhoefer, Felix and Winkler, Martin Wolfgang",
    title = "{Dirac Materials for Sub-MeV Dark Matter Detection: New Targets and Improved Formalism}",
    eprint = "1910.02091",
    archivePrefix = "arXiv",
    primaryClass = "hep-ph",
    reportNumber = "TTK-19-42",
    doi = "10.1103/PhysRevD.101.055005",
    journal = "Phys. Rev. D",
    volume = "101",
    number = "5",
    pages = "055005",
    year = "2020"
}

@article{Hochberg:2016ntt,
    author = "Hochberg, Yonit and Kahn, Yonatan and Lisanti, Mariangela and Tully, Christopher G. and Zurek, Kathryn M.",
    title = "{Directional detection of dark matter with two-dimensional targets}",
    eprint = "1606.08849",
    archivePrefix = "arXiv",
    primaryClass = "hep-ph",
    doi = "10.1016/j.physletb.2017.06.051",
    journal = "Phys. Lett. B",
    volume = "772",
    pages = "239--246",
    year = "2017"
}

@article{Cavoto:2017otc,
    author = "Cavoto, G. and Luchetta, F. and Polosa, A. D.",
    title = "{Sub-GeV Dark Matter Detection with Electron Recoils in Carbon Nanotubes}",
    eprint = "1706.02487",
    archivePrefix = "arXiv",
    primaryClass = "hep-ph",
    doi = "10.1016/j.physletb.2017.11.064",
    journal = "Phys. Lett. B",
    volume = "776",
    pages = "338--344",
    year = "2018"
}

@article{Griffin:2018bjn,
    author = "Griffin, Sinead and Knapen, Simon and Lin, Tongyan and Zurek, Kathryn M.",
    title = "{Directional Detection of Light Dark Matter with Polar Materials}",
    eprint = "1807.10291",
    archivePrefix = "arXiv",
    primaryClass = "hep-ph",
    doi = "10.1103/PhysRevD.98.115034",
    journal = "Phys. Rev. D",
    volume = "98",
    number = "11",
    pages = "115034",
    year = "2018"
}

@article{Derenzo:2016fse,
    author = "Derenzo, Stephen and Essig, Rouven and Massari, Andrea and Soto, Adr{\'\i}an and Yu, Tien-Tien",
    title = "{Direct Detection of sub-GeV Dark Matter with Scintillating Targets}",
    eprint = "1607.01009",
    archivePrefix = "arXiv",
    primaryClass = "hep-ph",
    doi = "10.1103/PhysRevD.96.016026",
    journal = "Phys. Rev. D",
    volume = "96",
    number = "1",
    pages = "016026",
    year = "2017"
}

@article{Das:2022srn,
    author = "Das, Anirban and Kurinsky, Noah and Leane, Rebecca K.",
    title = "{Dark Matter Induced Power in Quantum Devices}",
    eprint = "2210.09313",
    archivePrefix = "arXiv",
    primaryClass = "hep-ph",
    reportNumber = "SLAC-PUB-17691",
    doi = "10.1103/PhysRevLett.132.121801",
    journal = "Phys. Rev. Lett.",
    volume = "132",
    number = "12",
    pages = "121801",
    year = "2024"
}

@article{Das:2024jdz,
    author = "Das, Anirban and Kurinsky, Noah and Leane, Rebecca K.",
    title = "{Transmon Qubit constraints on dark matter-nucleon scattering}",
    eprint = "2405.00112",
    archivePrefix = "arXiv",
    primaryClass = "hep-ph",
    reportNumber = "SLAC-PUB-17769",
    doi = "10.1007/JHEP07(2024)233",
    journal = "JHEP",
    volume = "07",
    pages = "233",
    year = "2024"
}

@article{DAMIC-M:2025luv,
    author = "Aggarwal, K. and others",
    collaboration = "DAMIC-M",
    title = "{Probing Benchmark Models of Hidden-Sector Dark Matter with DAMIC-M}",
    eprint = "2503.14617",
    archivePrefix = "arXiv",
    primaryClass = "hep-ex",
    doi = "10.1103/2tcc-bqck",
    journal = "Phys. Rev. Lett.",
    volume = "135",
    number = "7",
    pages = "071002",
    year = "2025"
}

@inproceedings{Essig:2022dfa,
    author = "Essig, Rouven and Giovanetti, Graham K. and Kurinsky, Noah and McKinsey, Dan and Ramanathan, Karthik and Stifter, Kelly and Yu, Tien-Tien",
    title = "{Snowmass2021 Cosmic Frontier: The landscape of low-threshold dark matter direct detection in the next decade}",
    booktitle = "{2022 Snowmass Summer Study}",
    eprint = "2203.08297",
    archivePrefix = "arXiv",
    primaryClass = "hep-ph",
    reportNumber = "FERMILAB-CONF-22-181-PPD",
    month = "3",
    year = "2022"
}

@article{Emken:2017hnp,
    author = "Emken, Timon and Kouvaris, Chris and Nielsen, Niklas Gr\o{}nlund",
    title = "{The Sun as a sub-GeV Dark Matter Accelerator}",
    eprint = "1709.06573",
    archivePrefix = "arXiv",
    primaryClass = "hep-ph",
    reportNumber = "CP3-ORIGINS-2017-035",
    doi = "10.1103/PhysRevD.97.063007",
    journal = "Phys. Rev. D",
    volume = "97",
    number = "6",
    pages = "063007",
    year = "2018"
}

@article{Hochberg:2015fth,
    author = "Hochberg, Yonit and Pyle, Matt and Zhao, Yue and Zurek, Kathryn M.",
    title = "{Detecting Superlight Dark Matter with Fermi-Degenerate Materials}",
    eprint = "1512.04533",
    archivePrefix = "arXiv",
    primaryClass = "hep-ph",
    doi = "10.1007/JHEP08(2016)057",
    journal = "JHEP",
    volume = "08",
    pages = "057",
    year = "2016"
}

@article{Sanchez-Martinez:2019bac,
    author = "S\'anchez-Mart\'\i{}nez, Miguel-\'Angel and Robredo, I\~nigo and Bidauzarraga, Arkaitz and Bergara, Aitor and de Juan, Fernando and Grushin, Adolfo G. and Vergniory, Maia G.",
    title = "{Spectral and optical properties of Ag$_3$Au(Se$_2$,Te$_2$) and dark matter detection}",
    eprint = "1905.04805",
    archivePrefix = "arXiv",
    primaryClass = "cond-mat.mtrl-sci",
    doi = "10.1088/2515-7639/ab3ea2",
    journal = "Materials",
    volume = "3",
    pages = "014001",
    year = "2019"
}

@article{Hochberg:2022apz,
    author = "Hochberg, Yonit and Kahn, Yonatan F. and Leane, Rebecca K. and Rajendran, Surjeet and Van Tilburg, Ken and Yu, Tien-Tien and Zurek, Kathryn M.",
    title = "{New approaches to dark matter detection}",
    doi = "10.1038/s42254-022-00509-4",
    journal = "Nature Rev. Phys.",
    volume = "4",
    number = "10",
    pages = "637--641",
    year = "2022"
}

@article{XENON:2026qow,
    author = "Aprile, E. and others",
    collaboration = "XENON",
    title = "{Light Dark Matter Search with 7.8 Tonne-Year of Ionization-Only Data in XENONnT}",
    eprint = "2601.11296",
    archivePrefix = "arXiv",
    primaryClass = "hep-ex",
    month = "1",
    year = "2026"
}

@article{Cook:2024cgm,
    author = "Cook, Cameron and Blanco, Carlos and Smirnov, Juri",
    title = "{Deep learning optimal molecular scintillators for dark matter direct detection}",
    eprint = "2501.00091",
    archivePrefix = "arXiv",
    primaryClass = "hep-ph",
    month = "12",
    year = "2024"
}

@article{Simchony:2024kcn,
    author = "Simchony, Aviv and others",
    title = "{Diamond and SiC Detectors for Rare Event Searches}",
    doi = "10.1007/s10909-024-03148-4",
    journal = "J. Low Temp. Phys.",
    volume = "216",
    number = "1-2",
    pages = "363--370",
    year = "2024"
}

@article{Emken:2019tni,
    author = "Emken, Timon and Essig, Rouven and Kouvaris, Chris and Sholapurkar, Mukul",
    title = "{Direct Detection of Strongly Interacting Sub-GeV Dark Matter via Electron Recoils}",
    eprint = "1905.06348",
    archivePrefix = "arXiv",
    primaryClass = "hep-ph",
    reportNumber = "CERN-TH-2019-071, CP3-Origins-2019-18 DNRF90, YITP-SB-19-14",
    doi = "10.1088/1475-7516/2019/09/070",
    journal = "JCAP",
    volume = "09",
    pages = "070",
    year = "2019"
}

@article{Essig:2016crl,
    author = "Essig, Rouven and Mardon, Jeremy and Slone, Oren and Volansky, Tomer",
    title = "{Detection of sub-GeV Dark Matter and Solar Neutrinos via Chemical-Bond Breaking}",
    eprint = "1608.02940",
    archivePrefix = "arXiv",
    primaryClass = "hep-ph",
    doi = "10.1103/PhysRevD.95.056011",
    journal = "Phys. Rev. D",
    volume = "95",
    number = "5",
    pages = "056011",
    year = "2017"
}

@article{Griffin:2024cew,
    author = "Griffin, Sin\'ead M. and Hadas, Guy Daniel and Hochberg, Yonit and Inzani, Katherine and Lehmann, Benjamin V.",
    title = "{Dark Matter-Electron Detectors for Dark Matter-Nucleon Interactions}",
    eprint = "2412.16283",
    archivePrefix = "arXiv",
    primaryClass = "hep-ph",
    reportNumber = "MIT-CTP/5678",
    month = "12",
    year = "2024"
}

@article{Drukier:1986tm,
    author = "Drukier, A. K. and Freese, Katherine and Spergel, D. N.",
    title = "{Detecting Cold Dark Matter Candidates}",
    doi = "10.1103/PhysRevD.33.3495",
    journal = "Phys. Rev. D",
    volume = "33",
    pages = "3495--3508",
    year = "1986"
}

@article{Baxter:2019pnz,
    author = "Baxter, Daniel and Kahn, Yonatan and Krnjaic, Gordan",
    title = "{Electron Ionization via Dark Matter-Electron Scattering and the Migdal Effect}",
    eprint = "1908.00012",
    archivePrefix = "arXiv",
    primaryClass = "hep-ph",
    reportNumber = "FERMILAB-PUB-19-257-A",
    doi = "10.1103/PhysRevD.101.076014",
    journal = "Phys. Rev. D",
    volume = "101",
    number = "7",
    pages = "076014",
    year = "2020"
}

@article{Kurinsky:2020dpb,
    author = "Kurinsky, Noah and Baxter, Daniel and Kahn, Yonatan and Krnjaic, Gordan",
    title = "{Dark matter interpretation of excesses in multiple direct detection experiments}",
    eprint = "2002.06937",
    archivePrefix = "arXiv",
    primaryClass = "hep-ph",
    reportNumber = "FERMILAB-PUB-20-065-A",
    doi = "10.1103/PhysRevD.102.015017",
    journal = "Phys. Rev. D",
    volume = "102",
    number = "1",
    pages = "015017",
    year = "2020"
}

@article{Emken:2017erx,
    author = "Emken, Timon and Kouvaris, Chris and Shoemaker, Ian M.",
    title = "{Terrestrial Effects on Dark Matter-Electron Scattering Experiments}",
    eprint = "1702.07750",
    archivePrefix = "arXiv",
    primaryClass = "hep-ph",
    doi = "10.1103/PhysRevD.96.015018",
    journal = "Phys. Rev. D",
    volume = "96",
    number = "1",
    pages = "015018",
    year = "2017"
}

@article{Emken:2017qmp,
    author = "Emken, Timon and Kouvaris, Chris",
    title = "{DaMaSCUS: The Impact of Underground Scatterings on Direct Detection of Light Dark Matter}",
    eprint = "1706.02249",
    archivePrefix = "arXiv",
    primaryClass = "hep-ph",
    reportNumber = "CP3-ORIGINS-2017-20",
    doi = "10.1088/1475-7516/2017/10/031",
    journal = "JCAP",
    volume = "10",
    pages = "031",
    year = "2017"
}

@article{Catena:2019gfa,
    author = "Catena, Riccardo and Emken, Timon and Spaldin, Nicola A. and Tarantino, Walter",
    title = "{Atomic responses to general dark matter-electron interactions}",
    eprint = "1912.08204",
    archivePrefix = "arXiv",
    primaryClass = "hep-ph",
    doi = "10.1103/PhysRevResearch.2.033195",
    journal = "Phys. Rev. Res.",
    volume = "2",
    number = "3",
    pages = "033195",
    year = "2020",
    note = "[Erratum: Phys.Rev.Res. 7, 019001 (2025)]"
}

@article{Radick:2020qip,
    author = "Radick, Aria and Taki, Anna-Maria and Yu, Tien-Tien",
    title = "{Dependence of Dark Matter - Electron Scattering on the Galactic Dark Matter Velocity Distribution}",
    eprint = "2011.02493",
    archivePrefix = "arXiv",
    primaryClass = "hep-ph",
    doi = "10.1088/1475-7516/2021/02/004",
    journal = "JCAP",
    volume = "02",
    pages = "004",
    year = "2021"
}

@article{Gelmini:2020xir,
    author = "Gelmini, Graciela B. and Takhistov, Volodymyr and Vitagliano, Edoardo",
    title = "{Scalar direct detection: In-medium effects}",
    eprint = "2006.13909",
    archivePrefix = "arXiv",
    primaryClass = "hep-ph",
    doi = "10.1016/j.physletb.2020.135779",
    journal = "Phys. Lett. B",
    volume = "809",
    pages = "135779",
    year = "2020"
}

@article{Trickle:2020oki,
    author = "Trickle, Tanner and Zhang, Zhengkang and Zurek, Kathryn M.",
    title = "{Effective field theory of dark matter direct detection with collective excitations}",
    eprint = "2009.13534",
    archivePrefix = "arXiv",
    primaryClass = "hep-ph",
    reportNumber = "CALT-TH-2020-037",
    doi = "10.1103/PhysRevD.105.015001",
    journal = "Phys. Rev. D",
    volume = "105",
    number = "1",
    pages = "015001",
    year = "2022"
}

@article{Knapen:2021run,
    author = "Knapen, Simon and Kozaczuk, Jonathan and Lin, Tongyan",
    title = "{Dark matter-electron scattering in dielectrics}",
    eprint = "2101.08275",
    archivePrefix = "arXiv",
    primaryClass = "hep-ph",
    reportNumber = "CERN-TH-2021-013",
    doi = "10.1103/PhysRevD.104.015031",
    journal = "Phys. Rev. D",
    volume = "104",
    number = "1",
    pages = "015031",
    year = "2021"
}

@article{Hochberg:2025rjs,
    author = "Hochberg, Yonit and Khalaf, Majed and Lenoci, Alessandro and Ovadia, Rotem",
    title = "{Determining (All) Dark Matter-Electron Scattering Rates From Material Properties}",
    eprint = "2510.25835",
    archivePrefix = "arXiv",
    primaryClass = "hep-ph",
    month = "10",
    year = "2025"
}

@article{Berghaus:2026kmj,
    author = "Berghaus, Kim V. and Essig, Rouven and McDuffie, Megan H.",
    title = "{The Migdal effect in Semiconductors for the Effective Field Theory of Dark Matter Direct Detection}",
    eprint = "2603.13052",
    archivePrefix = "arXiv",
    primaryClass = "hep-ph",
    month = "3",
    year = "2026"
}

@article{Dreyer:2026bmz,
    author = "Dreyer, Cyrus and Essig, Rouven and Fernandez-Serra, Marivi and Hott, Megan and Singal, Aman",
    title = "{All-electron dark matter-electron scattering with random-phase approximation dielectric screening and local field effects}",
    eprint = "2603.12326",
    archivePrefix = "arXiv",
    primaryClass = "hep-ph",
    month = "3",
    year = "2026"
}

@article{Abbamonte:2025guf,
    author = "Abbamonte, P. and others",
    title = "{SPLENDOR: a novel detector platform to search for light dark matter with narrow-gap semiconductors}",
    eprint = "2507.17782",
    archivePrefix = "arXiv",
    primaryClass = "physics.ins-det",
    reportNumber = "LA-UR-25-26113",
    month = "7",
    year = "2025"
}

@article{Kavanagh:2017cru,
    author = "Kavanagh, Bradley J.",
    title = "{Earth scattering of superheavy dark matter: Updated constraints from detectors old and new}",
    eprint = "1712.04901",
    archivePrefix = "arXiv",
    primaryClass = "hep-ph",
    doi = "10.1103/PhysRevD.97.123013",
    journal = "Phys. Rev. D",
    volume = "97",
    number = "12",
    pages = "123013",
    year = "2018"
}

@article{Kavanagh:2016pyr,
    author = "Kavanagh, Bradley J. and Catena, Riccardo and Kouvaris, Chris",
    title = "{Signatures of Earth-scattering in the direct detection of Dark Matter}",
    eprint = "1611.05453",
    archivePrefix = "arXiv",
    primaryClass = "hep-ph",
    reportNumber = "CP3-ORIGINS-2016-050",
    doi = "10.1088/1475-7516/2017/01/012",
    journal = "JCAP",
    volume = "01",
    pages = "012",
    year = "2017"
}

@article{Sikivie:2002bj,
    author = "Sikivie, Pierre and Wick, Stuart",
    title = "{Solar wakes of dark matter flows}",
    eprint = "astro-ph/0203448",
    archivePrefix = "arXiv",
    doi = "10.1103/PhysRevD.66.023504",
    journal = "Phys. Rev. D",
    volume = "66",
    pages = "023504",
    year = "2002"
}

@article{Alenazi:2006wu,
    author = "Alenazi, Moqbil S. and Gondolo, Paolo",
    title = "{Phase-space distribution of unbound dark matter near the Sun}",
    eprint = "astro-ph/0608390",
    archivePrefix = "arXiv",
    doi = "10.1103/PhysRevD.74.083518",
    journal = "Phys. Rev. D",
    volume = "74",
    pages = "083518",
    year = "2006"
}

@article{Collar:1992qc,
    author = "Collar, J. I. and Avignone, F. T.",
    title = "{Diurnal modulation effects in cold dark matter experiments}",
    doi = "10.1016/0370-2693(92)90873-3",
    journal = "Phys. Lett. B",
    volume = "275",
    pages = "181--185",
    year = "1992"
}

@article{Collar:1993ss,
    author = "Collar, J. I. and Avignone, F. T.",
    title = "{The effect of elastic scattering in the Earth on cold dark matter experiments}",
    doi = "10.1103/PhysRevD.47.5238",
    journal = "Phys. Rev. D",
    volume = "47",
    pages = "5238--5246",
    year = "1993"
}

@article{Hasenbalg:1997hs,
    author = "Hasenbalg, F. and Abriola, D. and Avignone, F. T. and Collar, J. I. and Di Gregorio, D. E. and Gattone, A. O. and Huck, H. and Tomasi, D. and Urteaga, I.",
    title = "{Cold dark matter identification: Diurnal modulation revisited}",
    eprint = "astro-ph/9702165",
    archivePrefix = "arXiv",
    doi = "10.1103/PhysRevD.55.7350",
    journal = "Phys. Rev. D",
    volume = "55",
    pages = "7350--7355",
    year = "1997"
}

@article{Kouvaris:2015xga,
    author = "Kouvaris, Chris and Nielsen, Niklas Gr{\o}nlund",
    title = "{Daily modulation and gravitational focusing in direct dark matter search experiments}",
    eprint = "1505.02615",
    archivePrefix = "arXiv",
    primaryClass = "hep-ph",
    reportNumber = "CP3-ORIGINS-2015-015, DIAS-2015-15",
    doi = "10.1103/PhysRevD.92.075016",
    journal = "Phys. Rev. D",
    volume = "92",
    number = "7",
    pages = "075016",
    year = "2015"
}

@book{Cowan:1998ji,
    author = "Cowan, G.",
    title = "{Statistical data analysis}",
    isbn = "978-0-19-850156-5",
    year = "1998",
    publisher = "Oxford University Press"
}

@article{Leane:2025efj,
    author = "Leane, Rebecca K. and Beacom, John F.",
    title = "{Sub-GeV Dark Matter Direct Detection with Neutrino Observatories}",
    eprint = "2503.09685",
    archivePrefix = "arXiv",
    primaryClass = "hep-ph",
    reportNumber = "SLAC-PUB-250312",
    doi = "10.1103/mcyx-g4pd",
    journal = "Phys. Rev. Lett.",
    volume = "135",
    number = "19",
    pages = "191003",
    year = "2025"
}

@article{Santos-Olmsted:2025nuk,
    author = "Santos-Olmsted, Lillian and Leane, Rebecca K. and Blanco, Carlos and Beacom, John F.",
    title = "{Sub-GeV dark matter detection with dark rates in liquid scintillators}",
    eprint = "2512.13779",
    archivePrefix = "arXiv",
    primaryClass = "hep-ph",
    reportNumber = "SLAC-PUB-251210",
    doi = "10.1103/l4y7-5nkv",
    journal = "Phys. Rev. D",
    volume = "113",
    number = "10",
    pages = "103007",
    year = "2026"
}

@article{Kolmogorov:1933xxx,
  author       = {Kolmogorov, A. N.},
  title        = {Sulla determinazione empirica di una legge di distribuzione},
  journal      = {Giornale dell'Istituto Italiano degli Attuari},
  volume       = {4},
  pages        = {83--91},
  year         = {1933}
}

@article{Smirnov:1933xxx,
  author       = {Smirnov, N. V.},
  title        = {Estimate of deviation between empirical distribution functions in two independent samples},
  journal      = {Bulletin Moscow University},
  volume       = {2},
  pages        = {3--16},
  year         = {1933}
}

@article{Aaditya:2017xxx,
AUTHOR = {Ramdas, Aaditya and Trillos, Nicolás García and Cuturi, Marco},
TITLE = {On Wasserstein Two-Sample Testing and Related Families of Nonparametric Tests},
JOURNAL = {Entropy},
VOLUME = {19},
YEAR = {2017},
NUMBER = {2},
ARTICLE-NUMBER = {47},
URL = {https://www.mdpi.com/1099-4300/19/2/47},
ISSN = {1099-4300},
DOI = {10.3390/e19020047}
}

@article{Smith:2006ym,
    author = "Smith, Martin C. and others",
    title = "{The RAVE Survey: Constraining the Local Galactic Escape Speed}",
    eprint = "astro-ph/0611671",
    archivePrefix = "arXiv",
    doi = "10.1111/j.1365-2966.2007.11964.x",
    journal = "Mon. Not. Roy. Astron. Soc.",
    volume = "379",
    pages = "755--772",
    year = "2007"
}

@article{Bland:2016xxx,
   title={The Galaxy in Context: Structural, Kinematic, and Integrated Properties},
   volume={54},
   ISSN={1545-4282},
   url={http://dx.doi.org/10.1146/annurev-astro-081915-023441},
   DOI={10.1146/annurev-astro-081915-023441},
   number={1},
   journal={Annual Review of Astronomy and Astrophysics},
   publisher={Annual Reviews},
   author={Bland-Hawthorn, Joss and Gerhard, Ortwin},
   year={2016}, pages={529–596} }

@article{GRAVITY:2021xxx,
   title={Improved GRAVITY astrometric accuracy from modeling optical aberrations},
   volume={647},
   ISSN={1432-0746},
   url={http://dx.doi.org/10.1051/0004-6361/202040208},
   DOI={10.1051/0004-6361/202040208},
   journal={Astronomy and Astrophysics},
   publisher={EDP Sciences},
   author={Abuter, R. and others},
   year={2021},
   month=Mar, pages={A59} }

@ARTICLE{Schonrich:2010xxx,
       author = {{Sch{\"o}nrich}, Ralph and {Binney}, James and {Dehnen}, Walter},
        title = "{Local kinematics and the local standard of rest}",
      journal = {Monthly Notices of the Royal Astronomical Society},
         year = 2010,
        month = apr,
       volume = {403},
       number = {4},
        pages = {1829-1833},
          doi = {10.1111/j.1365-2966.2010.16253.x},
archivePrefix = {arXiv},
       eprint = {0912.3693},
 primaryClass = {astro-ph.GA},
       adsurl = {https://ui.adsabs.harvard.edu/abs/2010MNRAS.403.1829S}
}

@book{Fisher:1925xxx,
            year = {1925},
           title = {Statistical methods for research workers},
       publisher = {Oliver and Boyd},
          author = {Fisher, Ronald Aylmer},
             url = {https://repository.rothamsted.ac.uk/id/eprint/27823/}
}

@article{Mosteller:1948xxx,
author = {Frederick Mosteller},
title = {Questions and Answers},
journal = {The American Statistician},
volume = {2},
number = {5},
pages = {30--31},
year = {1948},
publisher = {Taylor \& Francis},
doi = {10.1080/00031305.1948.10483405},
URL = { 
    
    
        https://www.tandfonline.com/doi/abs/10.1080/00031305.1948.10483405
    

},}

\end{document}